\documentclass[aps,pra,twocolumn,superscriptaddress,floatfix]{revtex4-1}

\usepackage{bm}
\usepackage{amsmath}
\usepackage{latexsym}
\usepackage{amssymb}
\usepackage{graphicx}
\usepackage{color}
\usepackage{soul,xcolor}
\setstcolor{red}
\usepackage{ulem}
\usepackage{cancel}
\usepackage{tikz}
\usepackage[export]{adjustbox}
\usepackage{braket}
\usepackage{subfigure}

\usepackage[colorlinks=true, citecolor=blue, urlcolor=blue]{hyperref}\usepackage{epsf,graphics,graphicx}

\usepackage{float} 

\DeclareUnicodeCharacter{2009}{\,}

\newcommand{\beq}{\begin{equation}}
\newcommand{\eeq}{\end{equation}}

\def\bea{\begin{eqnarray}}
\def\eea{\end{eqnarray}}
\def\ba{\begin{array}}
\def\ea{\end{array}}

\usepackage{epsfig}

\input{epsf}

\begin{document}
\title{Chiral magnetic effect in lattice models of tilted multi-Weyl semimetals}
\author{Anirudha Menon} \email{amenon@ucdavis.edu} \affiliation{Department of Physics, University of California, Davis, California 95616, USA}
\author{Souvik Chattopadhay}	\email{souvikchatterjee812@gmail.com} \affiliation{Physics and Applied Mathematics Unit, Indian Statistical Institute, Kolkata 700108, India} 	 
\author{Banasri Basu}
	\email{banasri@isical.ac.in} \affiliation{Physics and Applied Mathematics Unit, Indian Statistical Institute, Kolkata 700108, India}

\date{\today}

\begin{abstract}
We study the chiral magnetic effect (CME) in tilted multi-Weyl Semimetals (WSM) employing a two-band lattice model. We focus on the type-II phase of  
mWSMs, introduced by incorporating a Lorentz symmetry violating tilt term. We add to the understanding of the CME and anomalous Hall effect (AHE) in the type-II phase of mWSMs and near the Lifshitz transition by varying tilt. Like the elementary WSM, our results also indicate that the Berry curvature drives the CME for higher monopole charges. We find a peak in the CME at the transition point and discuss its significance using the density of states. Along the way we also examine both observables as a function of the energy separation of the Weyl points.  
\end{abstract}

\maketitle


Weyl semimetals (WSM) realize a topologically nontrivial matter with low-energy electron excitations  described by gapless chiral fermions \cite{1d}( For reviews see \cite{rev1,rev2,5}). It is a topological state of matter possessing k-space singularities appearing at the touching point of the valence and conduction bands. In the prototypical WSM, a twofold band degeneracy at the Weyl point is broken linearly in momentum in all directions and the node is characterized by the topological winding number $n$. As compared to WSMs with $n = 1$, 
\cite{11,12,13}, $n$ can be generically greater than one in some materials \cite{14,cand12,16},  determined by the crystallographic point symmetries, and are called as multi WSMs (mWSMs). Recent theoretical reports claim $\textrm{SiSr}_2$ \cite{cand9} and $\textrm{HgCr}_2\textrm{Se}_4$ \cite{14, cand12} as possible candidates for mWSMs with monopole charge $n=2$. The double-Weyl (n = 2) and triple-Weyl (n = 3) semimetals have the quadratic and cubic energy dispersion relations, respectively. The dispersion anisotropy in mWSMs coupled with spin-momentum locking \cite{am8} has the potential to give rise to unique quantum effects and transport signatures \cite{huang17,new82,new83,new84,new86, AMBBTN,AMBB}.

Interestingly, large tilting of the Weyl cone, resulting to a Lifshitz transition, leads to a new class of materials called type-II WSMs \cite{WSMII,FP1,FP2}. The type-II WSM phase characterized by intriguing electronic transport properties due to a markedly different density of states at the Fermi level \cite{35,36,37,38,AMBB2018}. The existence of type-II WSM has been experimentally demonstrated \cite{32,33} while theoretical prediction shows that a type-II WSM can be engineered by applying strain or chemical doping to the original type-I WSM \cite{34}. Materials with this band structure have touching quasi-particle pocket Fermi surfaces at charge neutrality, compared to the point like Fermi surfaces for type-I WSMs. 

It is well known that the chiral anomaly plays an important role in the description of transport phenomena in WSMs. The chiral magnetic effect (CME) \cite{nielsen,CME0,CME1,CME2,CME3,CME4,CME5,CME6}, a significant transport phenomena, is realized in WSMs as well as in Dirac semimetals. In this effect, application of a magnetic field gives rise to a dissipation less electric current \cite{CME0} when the pair of Weyl nodes have different energies \cite{ZB1, new53, CME4, CME5,zwb2012}. Recent  experimental measurements have also confirmed a key signal of CME in Weyl/Dirac semimetals \cite{e1,e2,e3,e4,e5,e6,e7}. 

The existence of the chiral magnetic effect (CME) has been established within linear response theory by employing a two-band lattice model of WSMs \cite{CME1,CME2}. The topological properties of WSMs related to the static and dynamic CME has also been studied \cite{CME3}. However, the CME for the type-II phase of a generic mWSM has not been explored concretely; indeed, a systematic study of the transport properties of mWSMs with $tilt$ are scarce. In this paper, we try to provide a generalized framework for the study of tilt dependent transport properties, based on the linear response theory and using a realistic two-band lattice model of mWSM. We focus on the investigation of the role of tilt and the energy split between the Weyl nodes in the study of the CME. Additionally,  our investigation also reflects the dependence of these parameters on the anomalous Hall effect (AHE), in both the types of mWSMs. 

\section{The Lattice Model}

Motivated by the hint of the CME variation in the two mWSM phases in the minimal model (see appendix), we introduce a hybrid inversion and TRS broken two-band mWSM lattice model Hamiltonian for an in-depth investigation. This is constructed by adding a hopping integral term \cite{CME1, CME2}, which splits the node energy, to a generic TRS broken mWSM Hamiltonian \cite{Latticemod1, Latticemod2, AMBBTN}.

\beq \label{Ham}
H^n = d_0 + \vec{d}^n \cdot \vec{\sigma},
\eeq

where $\vec{\sigma} = [\sigma_x, \sigma_y, \sigma_z]$ are the vectorized Pauli matrices, and $d_0$ and $\vec{d}^n = [d_x^n, d_y^n, d_z^n]$ are lattice periodic functions of Bloch momenta, and $n$ is the postive integer monopole charge of the mWSM. The Hamiltonian is explicitly constructed based on the following parameters: $t_C$ represents tilt of the Weyl spectrum, $t_0$ represents the TRS breaking magnetization, $t_1$ controls the energy splitting between the Weyl nodes, and $t, t_z$ contribute to the Fermi velocity. With this information, we present the components of $D^n = [d_0, \vec{d}^n]$ for $n = 1, 2, 3$. In each case, the lattice constant has been set to unity and the Weyl points can be found at $(0, 0, \pm \pi/2)$.

\begin{widetext}
\begin{eqnarray}\label{Dcomp}
	D^1 &=& [t_C(\cos k_z + \cos k_x -1) + t_1 \sin k_z, t \sin k_x, t \sin k_y , t_z \cos k_z + t_0 (2- \cos k_x - \cos k_y)]  \nonumber \\
D^2 &=& [t_C(\cos k_z + \cos k_x -1) + t_1 \sin k_z, t (\cos k_x -\cos k_y), 2t \sin k_x \sin k_y, t_z \cos k_z \nonumber \\
&~& + t_0 (6 + \cos 2k_x + \cos 2k_y - 4\cos k_x - 4\cos k_y)]  \nonumber \\
D^3 &=& [t_C(\cos k_z + \cos k_x -1) + t_1 \sin k_z, t \sin k_x(1 - \cos k_x - 3(1 - \cos k_y)), \nonumber \\
&~& -t \sin k_y(1 - \cos k_y - 3(1 - \cos k_x)) , t_z \cos k_z + t_0 (6 + \cos 2k_x + \cos 2k_y - 4\cos k_x - 4\cos k_y)]  
\end{eqnarray}

\begin{eqnarray}\label{Egnval}
E_{\pm}^1 &=& t_C(\cos k_z + \cos k_x -1) + t_1 \sin k_z \pm  \{ t^2 \sin^2 k_x + t^2\sin^2 k_y +[t_z \cos k_z + t_0 (2- \cos k_x - \cos k_y)]^2\}^{1/2}
\nonumber \\ E_{\pm}^2 &=& t_C(\cos k_z + \cos k_x -1) + t_1 \sin k_z \nonumber \\ &~& \pm \{t^2(\cos k_x -\cos 
k_y)^2 + 4t^2 \sin^2 k_y \sin^2 k_y + [t_z \cos k_z + t_0 (6 + \cos 2k_x
+ \cos 2k_y - 4\cos k_x - 4\cos k_y)]^2\}^{1/2} \nonumber \\
E_{\pm}^3 &=& t_C(\cos k_z + \cos k_x -1) + t_1 \sin k_z  \pm  \{ t^2 (2 + \cos 
k_x - 3 \cos k_y)^2 \sin^2 k_x + (2 + \cos k_y - 3 \cos k_x )^2 \sin^2
k_y + \nonumber \\ &~& \hspace{180pt} [t_z \cos k_z + t_0 (6 + \cos 2k_x + \cos 2k_y - 4\cos k_x -
4\cos k_y)]^2\}^{1/2} \nonumber \\
\end{eqnarray}

\end{widetext}

\begin{figure}
\centering
\includegraphics[scale=.33]{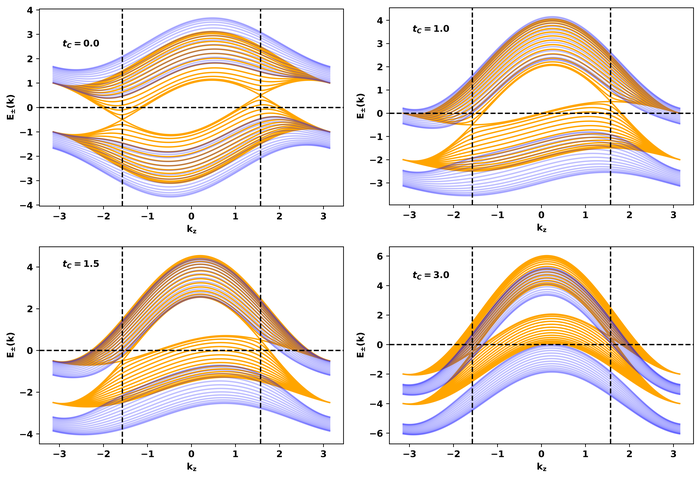}
\caption{The dispersion relations for $n=1$ mWSMs are presented at four values of tilt $t_C$. $t_C = 0$ corresponds to the type-I phase, $t_C = 1$ corresponds to the Lifshitz transition point, and $t_C = 1.5, 3$ represent the type-II phase. The orange bands correspond to $k_x = 0$ and the blue bands correspond to $k_x = 1$. The Weyl nodes are located at $k_x = 0, k_y = 0, k_z = \pm \pi/2$ (meeting point of orange bands), and $t_1 = 0.5$ sets the energy difference between them. Parameters for the plot are: $t_0 = 1, t_z = 1, t = 1$.}
\label{FigsE1}
\end{figure}

\begin{figure}
\centering
\includegraphics[scale=.33]{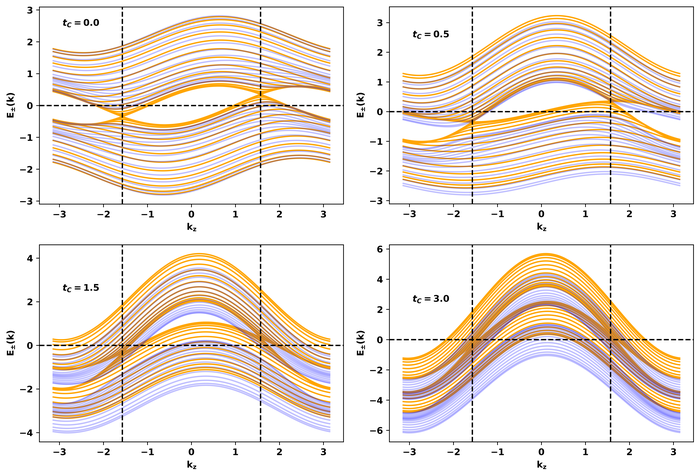}
\caption{The dispersion relations for $n=2$ mWSMs are presented at four values of tilt $t_C$. $t_C = 0$ corresponds to the type-I phase, $t_C = 0.5$ corresponds to the Lifshitz transition point, and $t_C = 1.5, 3$ represent the type-II phase. The orange bands correspond to $k_x = 0$ and the blue bands correspond to $k_x = 1$. The Weyl nodes are located at $k_x = 0, k_y = 0, k_z = \pm \pi/2$ (meeting point of orange bands), and $t_1 = 0.35$ sets the energy difference between them. Parameters for the plot are: $t_0 = 0.25, t_z = 0.5, t = 0.5$.}
\label{FigsE2}
\end{figure}

\begin{figure}
\centering
\includegraphics[scale=.33]{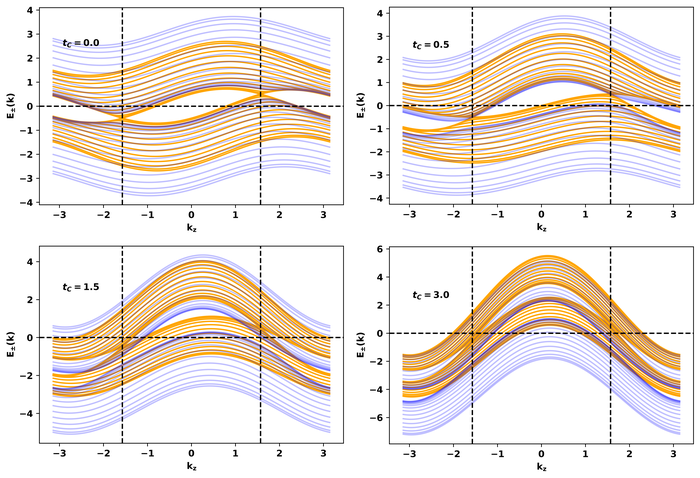}
\caption{The dispersion relations for $n=3$ mWSMs are presented at four values of tilt $t_C$. $t_C = 0$ corresponds to the type-I phase, $t_C = 0.5$ corresponds to the Lifshitz transition point, and $t_C = 1.5, 3$ represent the type-II phase. The orange bands correspond to $k_x = 0$ and the blue bands correspond to $k_x = 1$. The Weyl nodes are located at $k_x = 0, k_y = 0, k_z = \pm \pi/2$ (meeting point of orange bands), and $t_1 = 0.5$ sets the energy difference between them. Parameters for the plot are: $t_0 = 0.25, t_z = 0.5, t = 0.5$.}
\label{FigsE3}
\end{figure}

The details of the energy dispersion for this class of models is shown in Eqn.(\ref{Egnval}) and plotted in Figs. \ref{FigsE1}, \ref{FigsE2}, \& \ref{FigsE3}. For each value of monopole charge, we consider four values of tilt corresponding to the type-I phase, the Lifshitz transition point, the type-II phase, and the large tilt case. We wish to draw the reader's attention to the crucial difference between the two mWSM phases, namely, that the the type-II phase hosts both electron and hole pockets at the same Weyl point.

\section{The Chiral Magnetic Effect}

The chiral magnetic effect is the phenomenon of a dissipation-less electric current {\bf J} generated by and in the same direction of an applied magnetic field {\bf B}. The expression for the chiral magnetic effect (CME) coefficient $\alpha$, defined as ${\bf J} = \alpha {\bf B} $, can be calculated in the linear response regime using the antisymmetric part of the current-current correlation function as $\Pi^{ij}_{\textrm{anti}} ({\bf q}, \omega) = i\alpha ({\bf q}, \omega) \epsilon^{ijk}q_k $. In the uniform limit, following \cite{CME1,new53}, the CME coefficient is given by

\begin{eqnarray}\label{cmef0}
\alpha^i &=& \frac{e^2}{\hbar} \int \frac{d^3k}{(2\pi)^3} \sum_{t = \pm} \left ( \frac{v_{{\bf k},+} + v_{{\bf k},-} }{2} \cdot \Omega_t ({\bf k}) f_t ({\bf k}) \right. \nonumber \\ &~& \left. - d({\bf k}) \frac{v_{{\bf k},t}  \cdot \Omega_t ({\bf k}) - v_{{\bf k},t}^i \Omega_t^i ({\bf k})}{2} \frac{\partial f_t ({\bf k})}{\partial E_t} \right),
\end{eqnarray}

with $i = x, y, z$. In the expressions above, $\Omega_{\pm}^i ({\bf k}) = \pm \epsilon^{ijl} \frac{1}{4d^3({\bf k})} {\bf d}({\bf k}) \cdot (\frac{\partial {\bf d}({\bf k})}{d k_j} - \frac{\partial {\bf d}({\bf k})}{d k_l} )$ is the Berry curvature, $f({\bf k})$ is the Fermi-Dirac distribution (with $\mu$ as chemical potential), and $v_{{\bf k},\pm} = \frac{1}{\hbar} \nabla_{\bf k} E_{\pm} ({\bf k})$. As argued in \cite{new53}, $\alpha^i$ depends very weakly on the choice of direction. Hence, we consider the spatially averaged CME coefficient $\alpha = \frac{\alpha_x + \alpha_y + \alpha_z}{3}$ in what follows.

\begin{eqnarray}\label{cmef}
\alpha &=& \frac{e^2}{\hbar} \int \frac{d^3k}{(2\pi)^3} \sum_{t = \pm} \left ( \frac{v_{{\bf k},+} + v_{{\bf k},-} }{2} \cdot \Omega_t ({\bf k}) f_t ({\bf k}) \right. \nonumber \\ &~& \left. -\frac{1}{3} d({\bf k}) v_{{\bf k},t}  \cdot \Omega_t ({\bf k}) \frac{\partial f_t ({\bf k})}{\partial E_t} \right)
\end{eqnarray}

Employing equation (\ref{cmef}) we calculate the CME coefficient for our model Hamiltonian. All calculations are done for system sizes of $400^3$ or $500^3$ sites to avoid finite size effects, with the exception of the density of states (DOS). The next three plots are the central results of this work. As witnessed in \cite{CME1,CME2}, the overall sign of both quantities can be model dependent, and to keep this discussion simple, we consider the absolute values of $\alpha$.


\begin{figure}
\centering
\subfigure[]{\includegraphics[scale=.55]{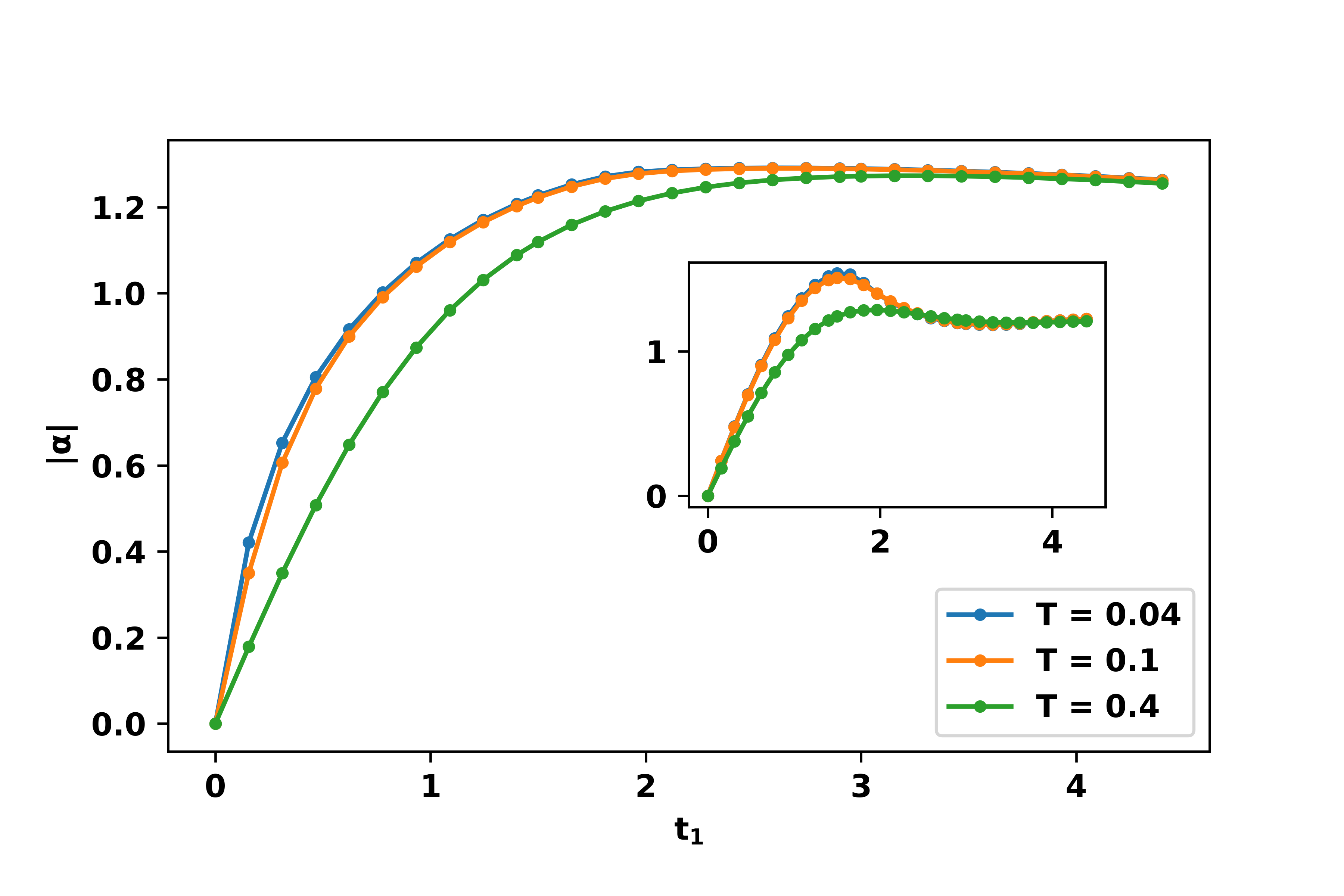}}
\subfigure[]{\includegraphics[scale=.55]{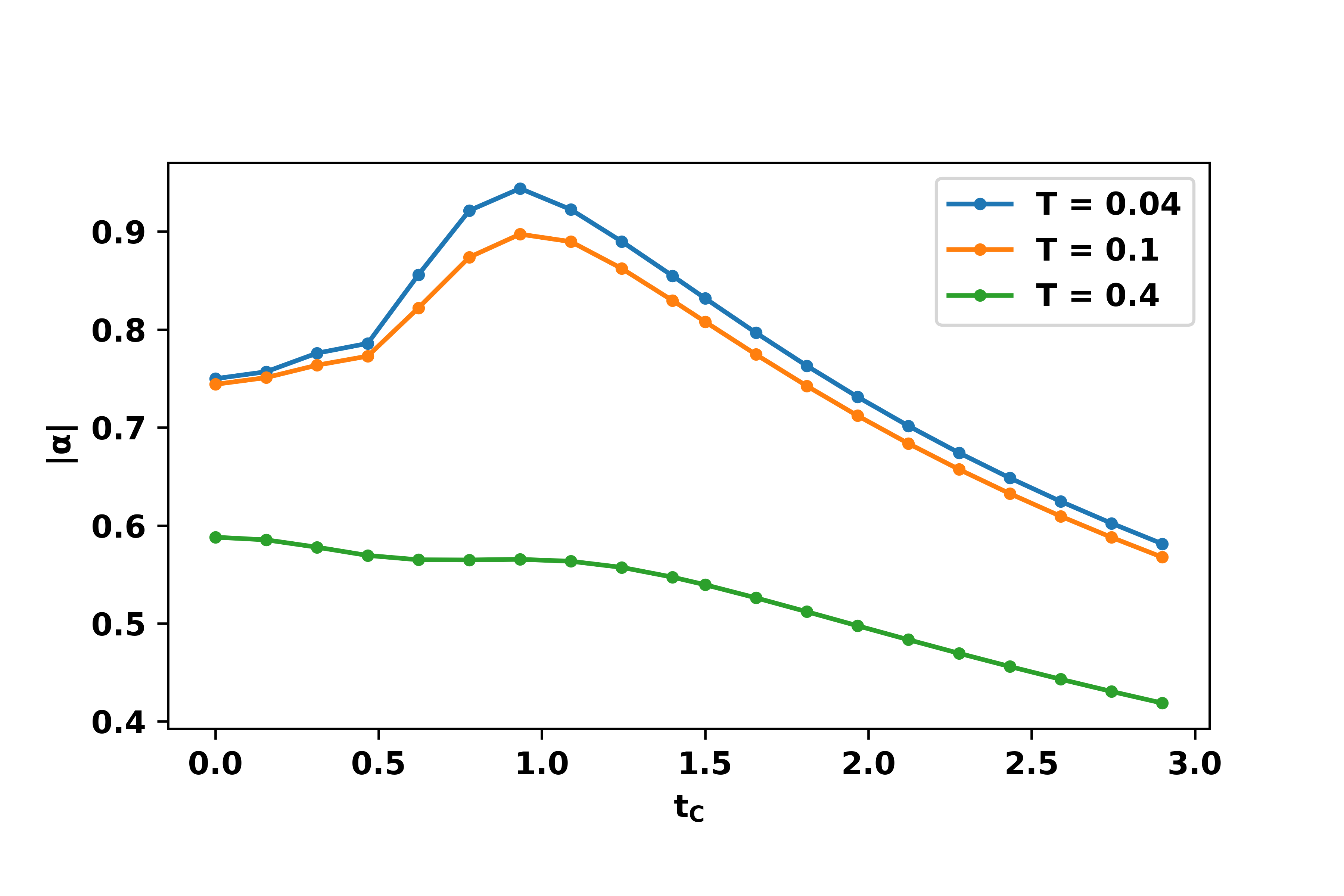}}
\caption{(a) The variation of $|\alpha|$ is plotted (in units of $e^2/h^2$) with $t_1$ for $n=1$ mWSMs, with the type-II phase shown in the mainframe ($t_C = 1.5$) and the type-I phase in the inset ($t_C = 0$). (b) The variation of $|\alpha|$ is plotted with $t_C$ for $n=1$ mWSMs with ($t_1 = 0.5$). Parameters common to both plots are: $t_0 = 1, \mu = 0, t_z = 1, t = 1$.}
\label{Figs1}
\end{figure}

\begin{figure}
\centering
\includegraphics[scale=.55]{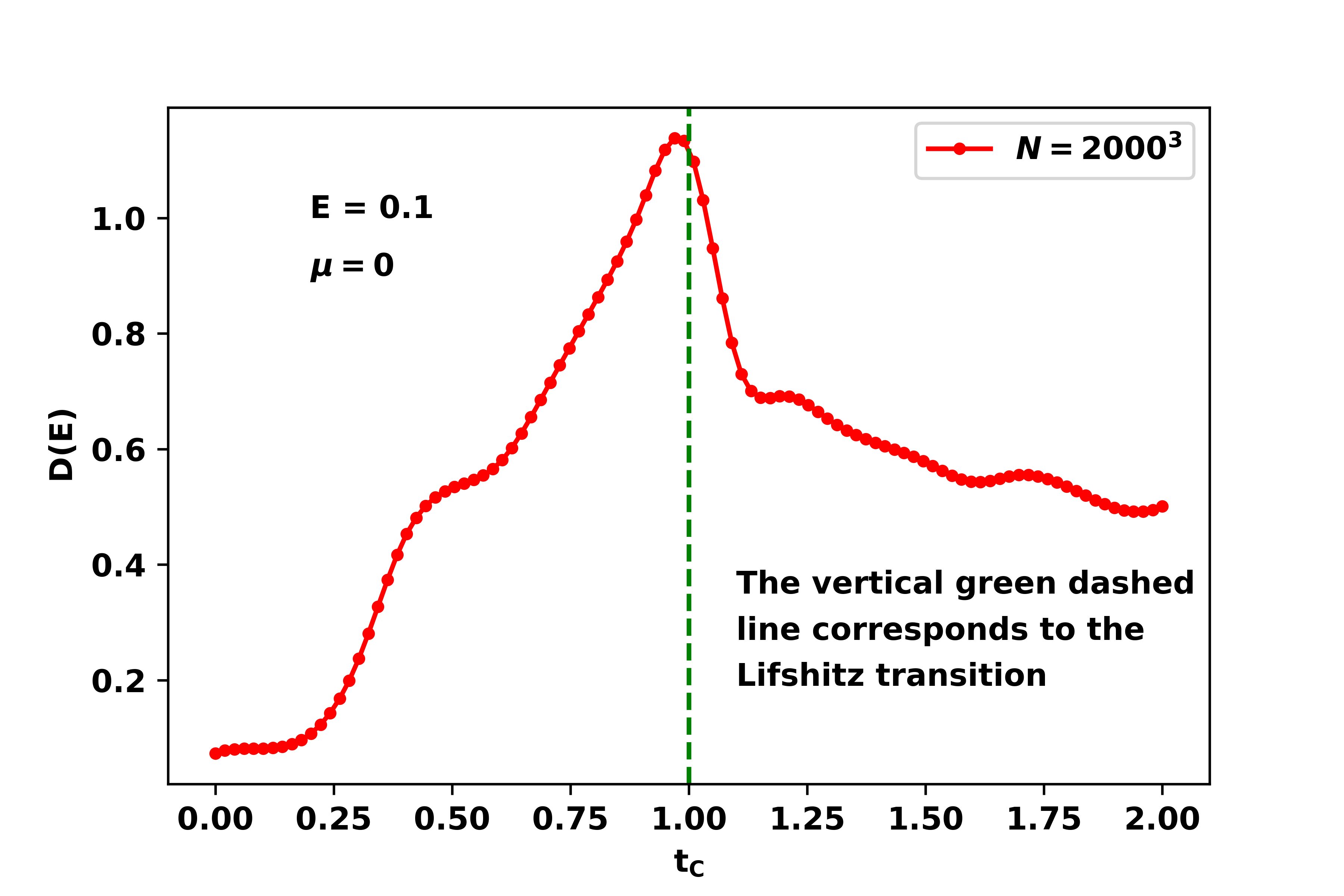}
\caption{The variation of the density of states (DOS) is plotted versus $t_C$ for $n=1$ mWSMs. The data converges for a lattice sizing of $2000^3$ sites. The peak in the DOS corresponds to the Lifshitz transition at $t_C = 1$. Parameters for the plot are: $t_0 = 1, \mu = 0, t_z = 1, t = 1$, $t_1 = 0.5$.}
\label{FigsDOS}
\end{figure}

Fig. \ref{Figs1} (a) shows the variation of the chiral magnetic parameter $\alpha$ with $t_1$ for $n=1$ mWSMs at three different temperatures. The type-I results, which have been independently established by other authors \cite{CME1}, are presented in the inset. Both phases of the WSM show an approximately constant $\alpha$ for values of $t_1 > 3$ at all temperatures. In Fig. \ref{Figs1} (b) we plot the CME parameter as a function of the tilt for three different temperatures. The type-I to type-II Lifshitz transition occurs at $t_C = 1$. At lower temperatures, one observes a possible signature of such a transition characterized by the orange and blue peaks. $\alpha$ shows an increasing trend leading up to the Lifshitz transition and decreases subsequently. The peak ceases to exist at a higher temperature ($T = 0.4$), likely due to thermalization as observed from the green line.

The CME in mWSMs is generated by the Berry curvature despite the fact that $\Omega_t(\bf k)$ is independent of $t_1$, due to its coupling to the velocity in Eqn.(\ref{cmef}). The integrand of this equation is an exponentially decreasing function of $t_1$ which may be determined from the Fermi-Dirac distribution. So, the CME integral should approach zero asymptotically for $t_1 \rightarrow \infty$. This is not reflected in the parameter range we have chosen, where $|\alpha|$ appears to be constant for $t_1 > 3$. The dependence of $\alpha$ on $t_C$ is similar, and in this case the asymptotic effects set in faster as seen in figure \ref{Figs1} (b).

\begin{figure}
\centering
\subfigure[]{\includegraphics[scale=.55]{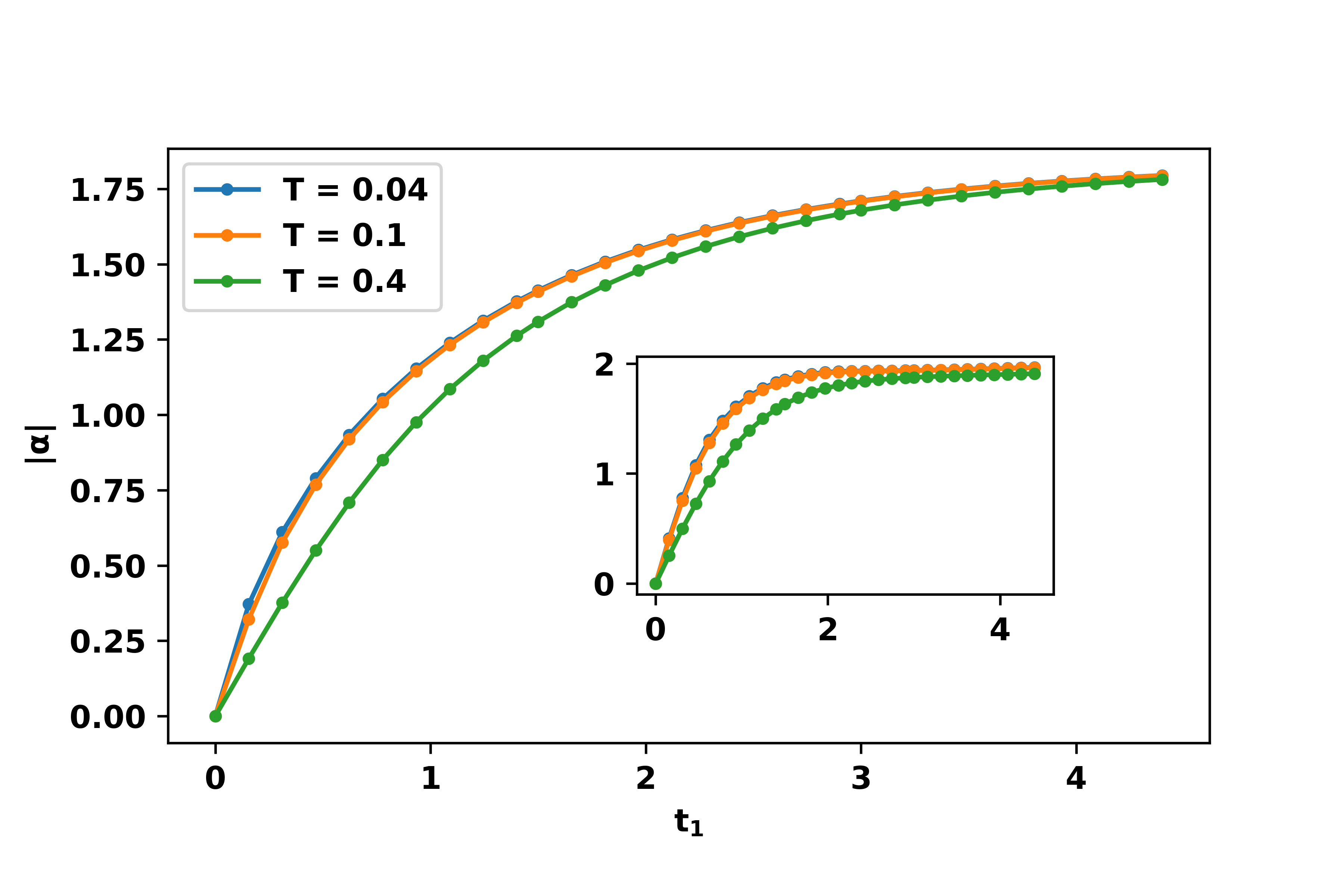}}
\subfigure[]{\includegraphics[scale=.55]{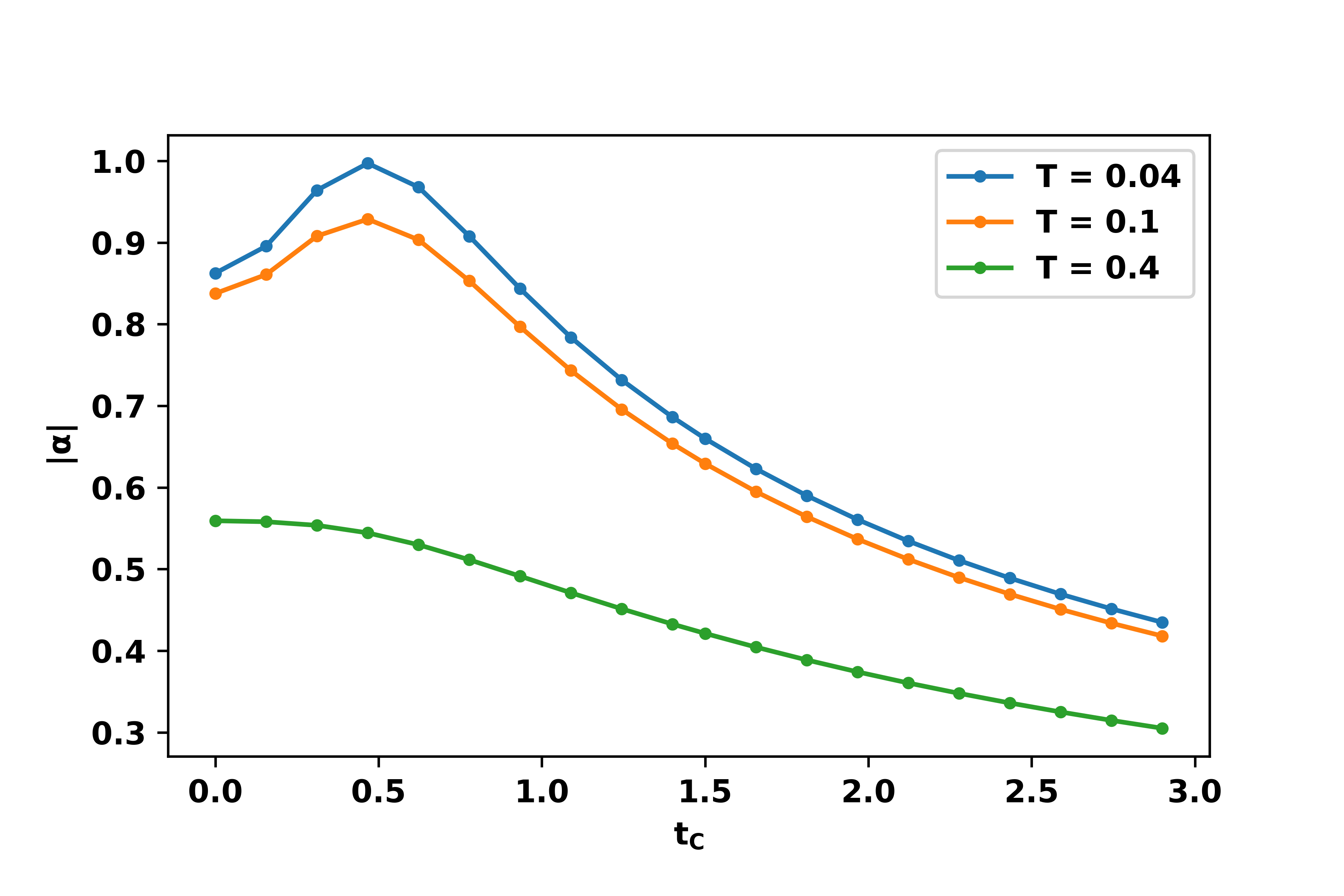}}
\caption{(a) The variation of $|\alpha|$ is plotted (in units of $e^2/h^2$) with $t_1$ for $n=2$ mWSMs, with the type-II phase shown in the mainframe ($t_C = 1.5$) and the type-I phase in the inset ($t_C = 0$). (b) The variation of $|\alpha|$ is plotted with $t_C$ for $n=2$ mWSMs ($t_1 = 0.35$). Parameters common to both plots are: $t_0 = 0.25, \mu = 0, t_z = 0.5, t = 0.5$.}
\label{Figs2}
\end{figure}

Since the Berry curvature is independent of $t_1$ and $t_C$ by construction, a question arises as to the origin of the peaks in Fig. \ref{Figs1} (b). Typically, conduction is influenced by carrier concentration and a useful diagnostic for this is the density of states. To this end, we examine the DOS $D(E)$ of the system defined as

\begin{equation} \label{DOS}
D^n(E) \equiv \sum_{s = \pm} \int_{\mathbb{R}^3} \frac{d^3 k}{(2\pi)^3} \delta [E - E^n_s(k)].
\end{equation}

$D^n(E)$ is plotted as a function of $t_C$ in Fig. \ref{FigsDOS} for $n=1$ mWSM at $E=0.1$s, i.e., close to the Fermi surface. Unlike the other observables computed in this manuscript, the DOS data converges for $2000^3$ lattice sites making it a computational challenge. We observe a sharp peak in the density of states at the Lifshitz transition $t_C = 1$ concurrent with the CME peaks in Fig. \ref{Figs1} (b). The DOS is temperature independent, which explains why the CME peaks are preserved in the low temperature regime.

\begin{figure}
\centering
\subfigure[]{\includegraphics[scale=.55]{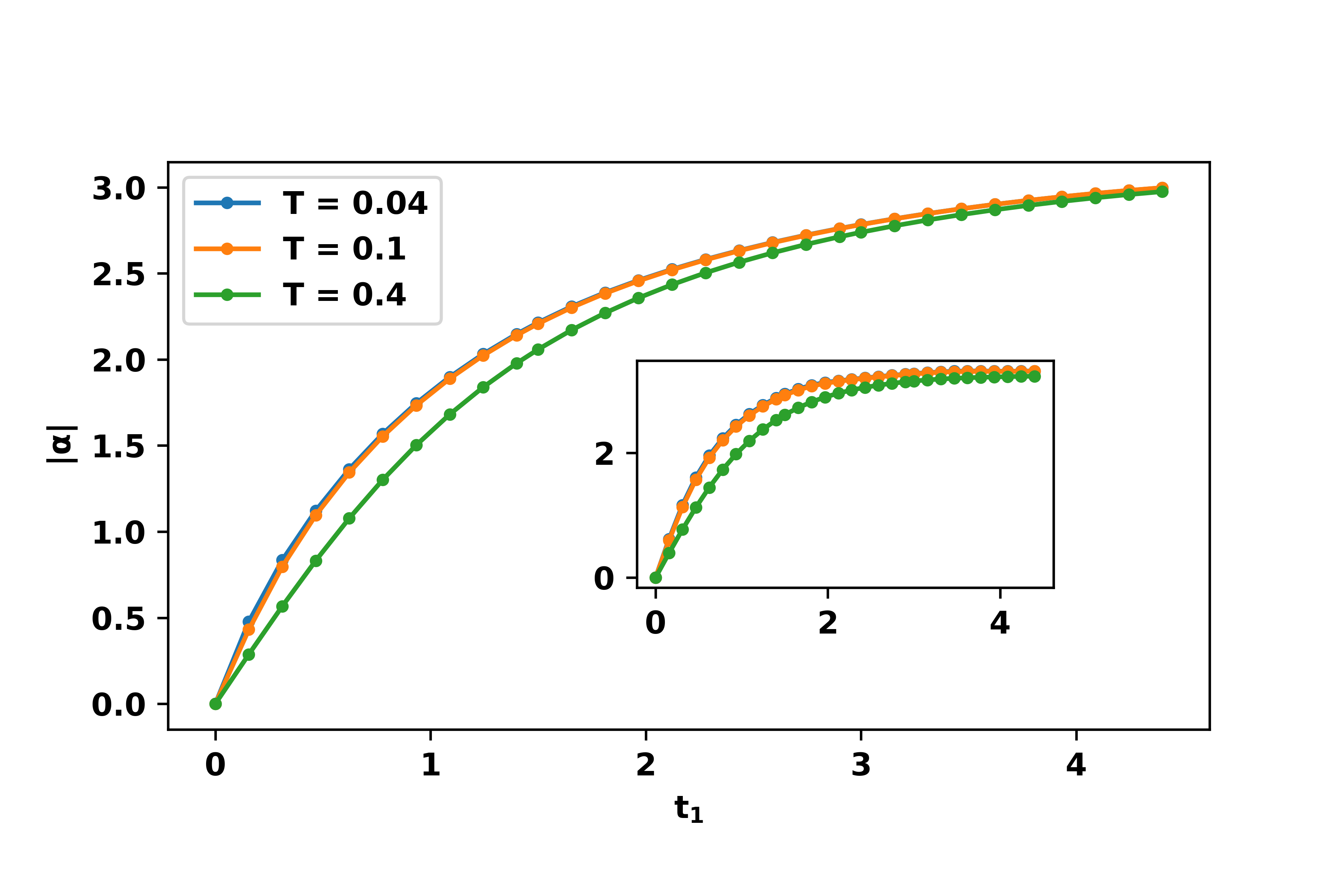}}
\subfigure[]{\includegraphics[scale=.55]{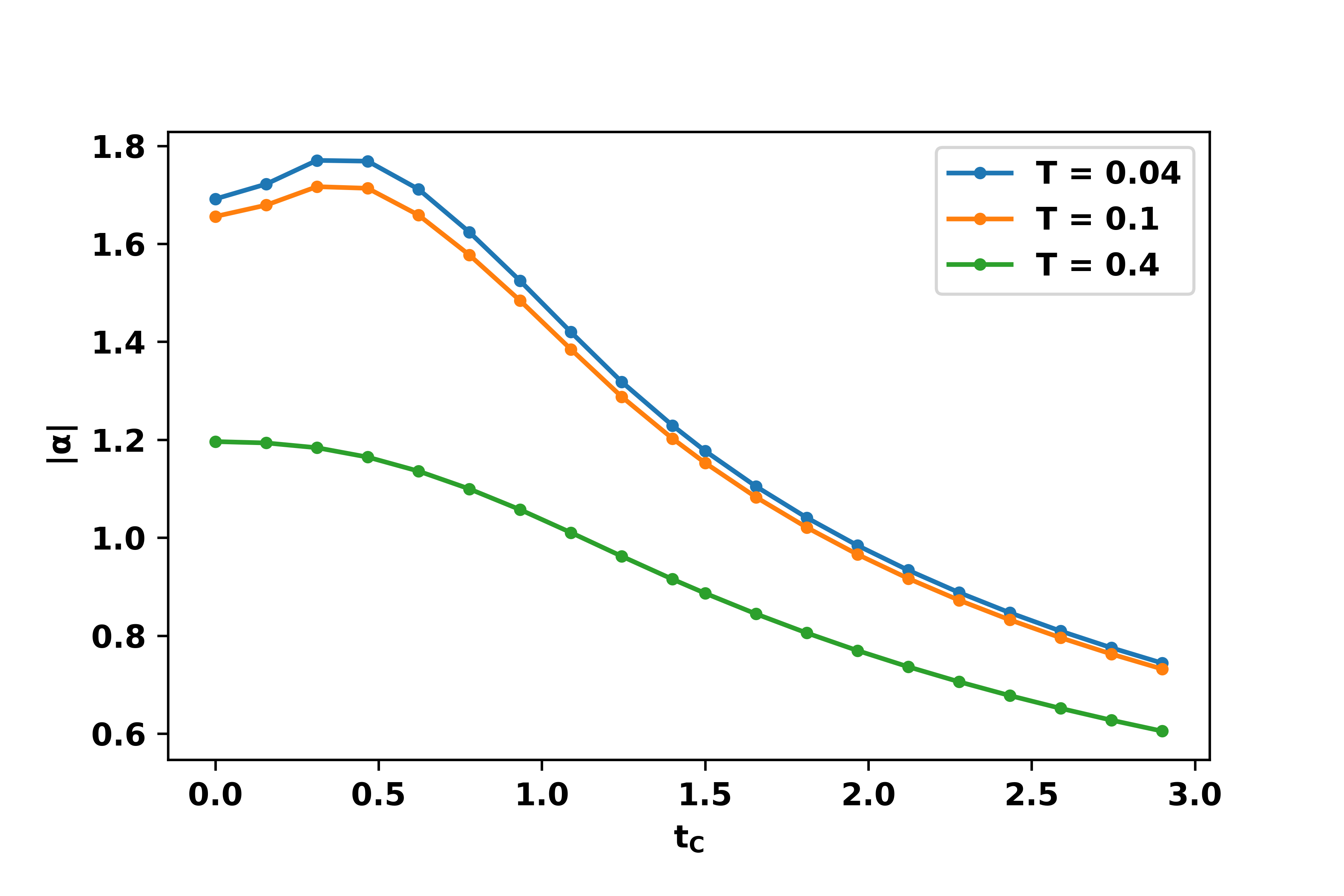}}
\caption{(a) The variation of $|\alpha|$ is plotted (in units of $e^2/h^2$) with $t_1$ for $n=3$ mWSMs, with the type-II phase shown in the mainframe ($t_C = 1.5$) and the type-I phase in the inset ($t_C = 0$). (b) The variation of $|\alpha|$ is plotted with $t_C$ for $n=3$ mWSMs ($t_1 = 0.5$). Parameters common to both plots are: $t_0 = 0.25, \mu = 0, t_z = 0.5, t = 0.5$.}
\label{Figs3}
\end{figure}

Examining the plots in Figs. \ref{Figs2} and \ref{Figs3} we find that the general observations made about the $n=1$ mWSMs continue to hold rather well in the $n=2$ and $n=3$ mWSM phases. A further point to illustrate that the possibility that the peak in the $n=1$ type-I phase CME (as a function of $t_1$) is parameter dependent, is its absence in the $n=2,3$ cases. In contrast, the peak reflecting the Lifshitz transition is preserved in the higher monopole charge cases, as can be seen from Figs. \ref{Figs2} (b) and \ref{Figs3} (b), where the transition occurs at $t_C = 0.5$. Such a signature which is independent of monopole charge can be experimentally verified as discussed later and can serve as a handle for phase characterization. The $D^n(E)$ data is a computational challenge for $n=2$ and $n=3$ mWSMs, and we don't pursue it in this work.

\section{The Quantum Anomalous Hall phase} 

\begin{figure}
\centering
\subfigure[]{\includegraphics[scale=.55]{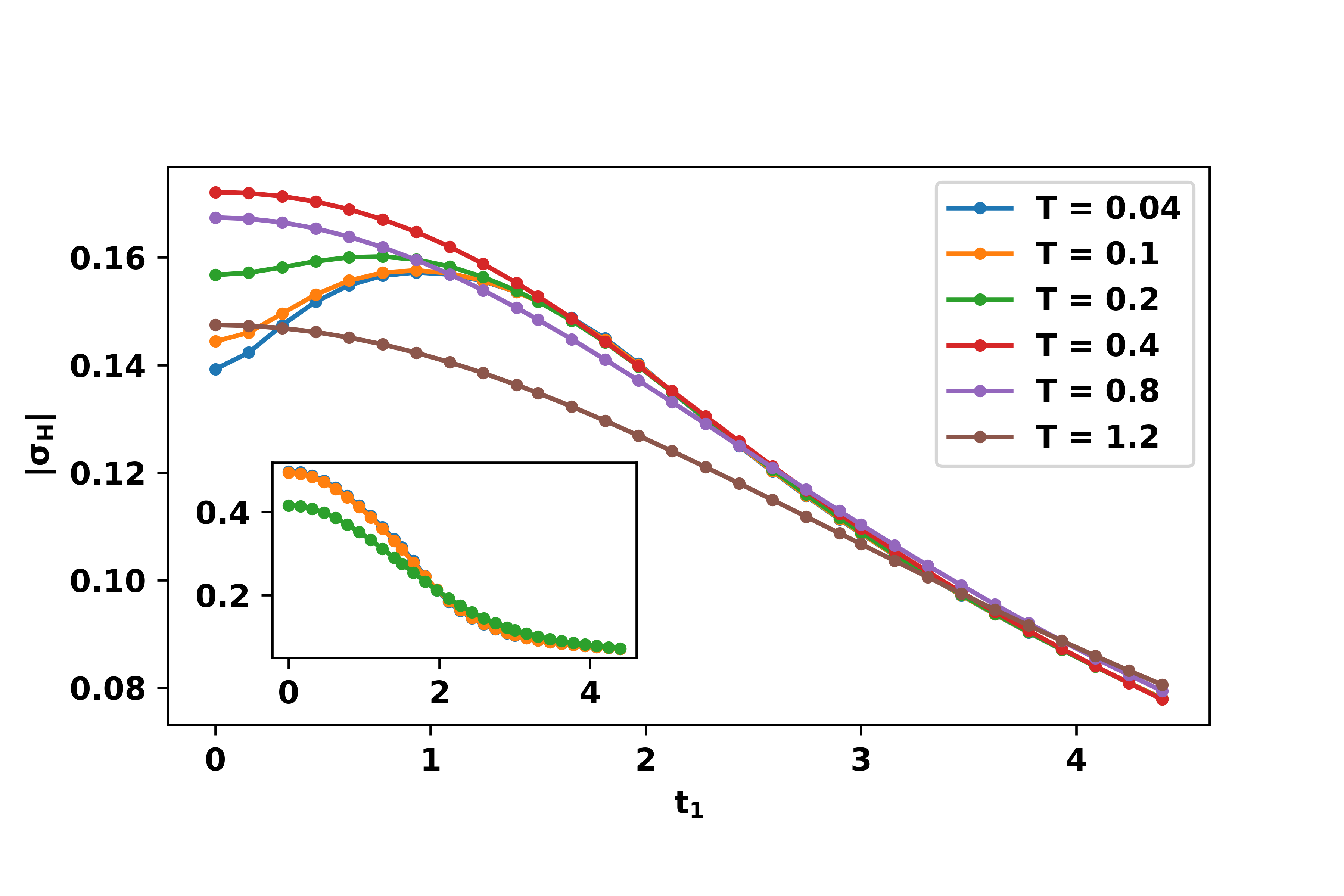}}
\subfigure[]{\includegraphics[scale=.55]{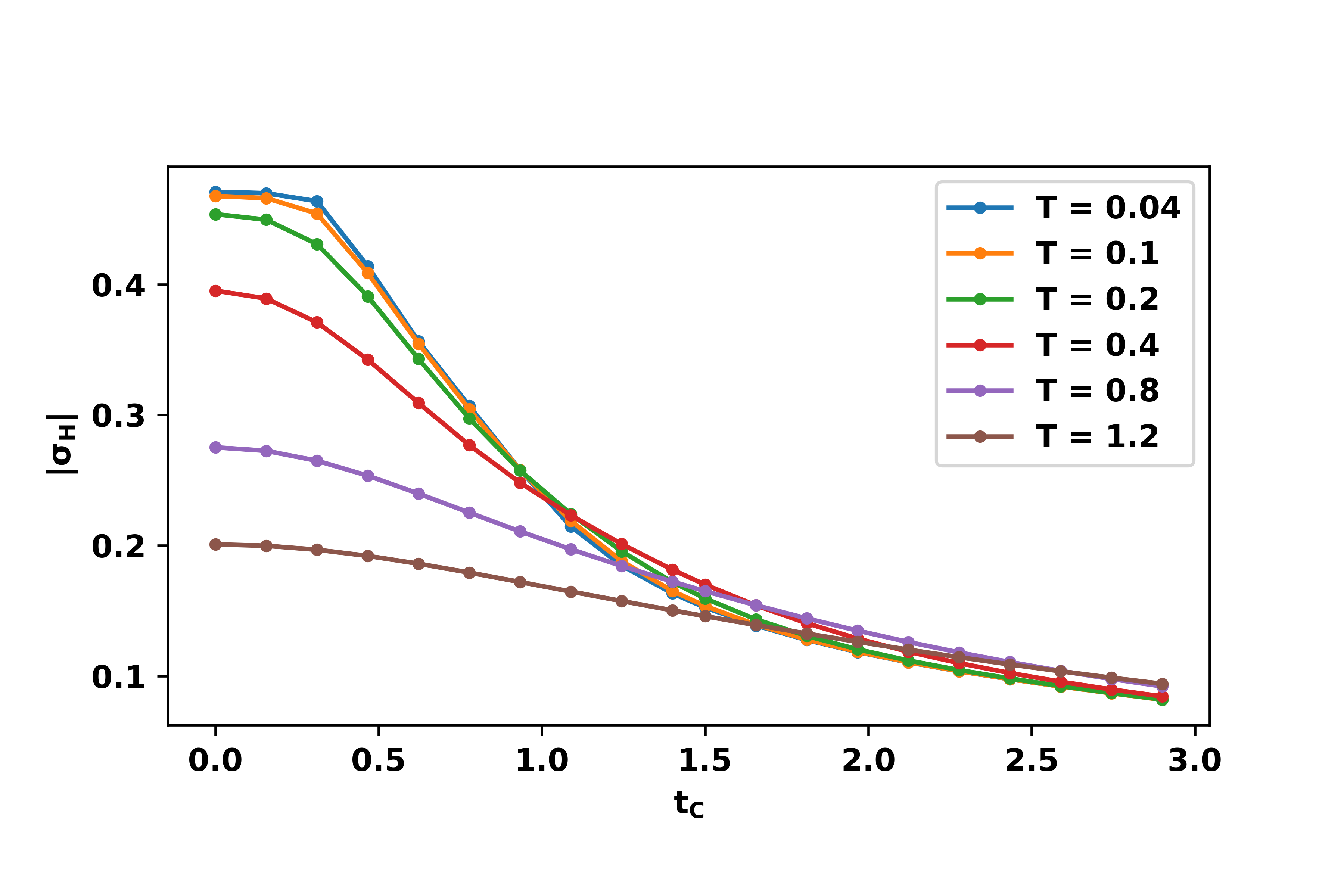}}
\subfigure[]{\includegraphics[scale=.55]{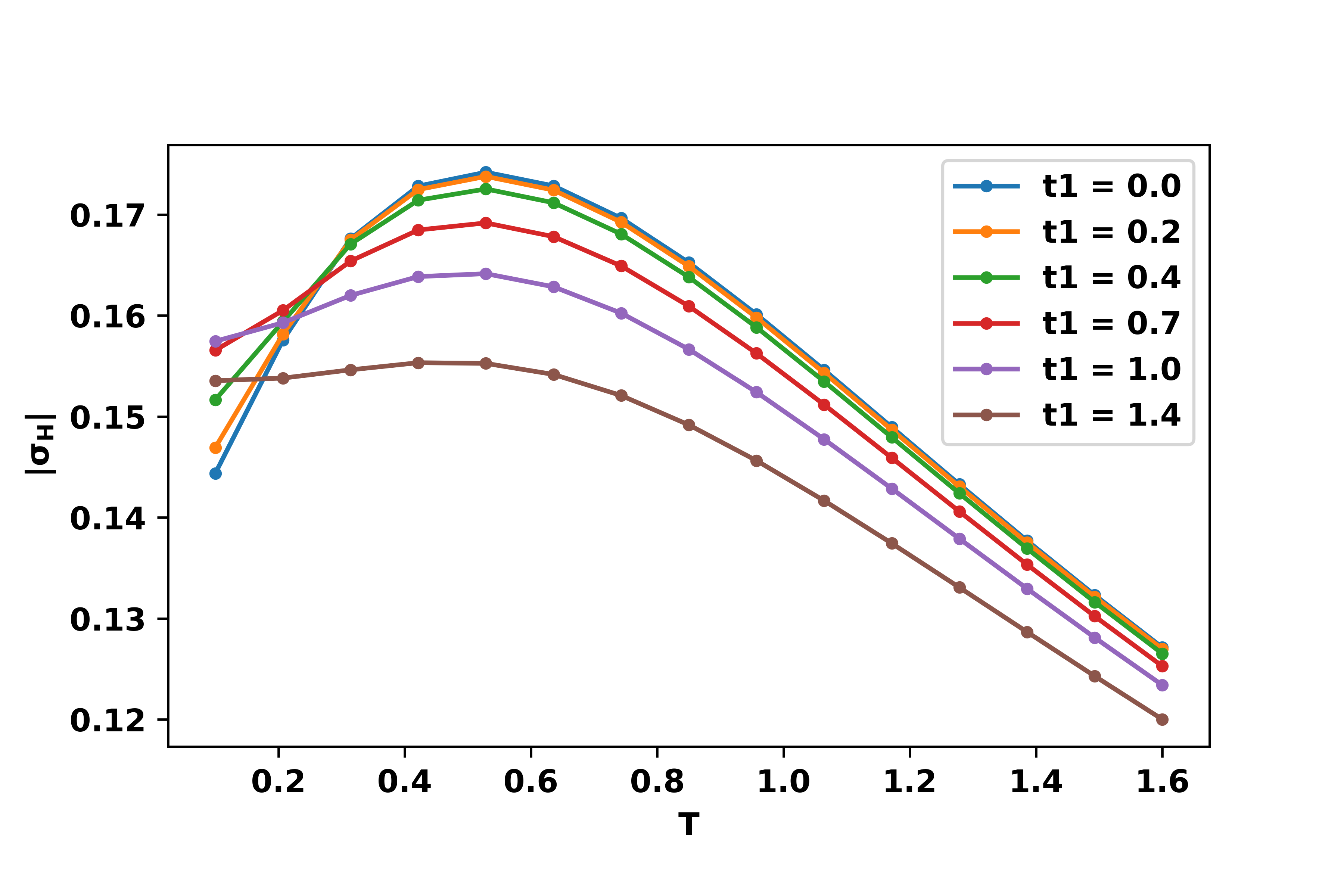}}
\caption{(a) The variation of $|\sigma_{H}|$ is plotted (in units of $e^2/h$) with $t_1$ for $n=1$ mWSMs, with the type-II phase shown in the mainframe ($t_C = 1.5$) and the type-I phase in the inset ($t_C = 0$). (b) The variation of $|\sigma_{H}|$ is plotted with $t_C$ for $n=1$ mWSMs ($t_1 = 0.5$). (c) The variation of $|\sigma_H|$ is plotted as a function of temperature for different values of $t_1$ for the $n=1$ mWSM with $t_C = 1.5$. For small values of $t_1$, the AHE shows a peak as a function of temperature. Parameters common to all plots are: $t_0 = 1, \mu = 0, t_z = 1, t = 1$.}
\label{Figs4}
\end{figure}

The quantum anomalous Hall phase of the TRS broken WSM is a hallmark effect \cite{ah1,ZB1,new53}. Compared to single WSMs, the transport properties of mWSMs  are modified by higher monopole charges or winding numbers resulting modification in anomalous Hall conductivity.   
 To study the QAH phase of type-I and type-II mWSMs, we may write the  anomalous Hall conductivity  in terms  of the Berry curvature as 

\begin{equation}\label{halle}
\sigma_H = \frac{e^2}{\hbar} \int \frac{d^3k}{(2\pi)^3} \sum_{t = \pm} \Omega_t^z ({\bf k}) f_t ({\bf k})
\end{equation}
With our model lattice Hamiltonian of mWSM we study the variation of $\sigma_H$ with respect to different model parameters. The anomalous Hall effect (AHE) for the $n=1$ mWSM or simple WSM is plotted in Fig. \ref{Figs4} (a) as a function of $t_1$. The type-I phase of mWSMs in the limit of charge neutrality should reach the well-known vacuum anomalous Hall conductivity given by $|\sigma_H| = n\frac{e^2 Q}{2\pi^2 \hbar}$ ($2Q = \pi$ is the node separation) which is indeed seen in the inset at $t_1 = 0$. 

\begin{figure}
\centering
\includegraphics[scale=0.7]{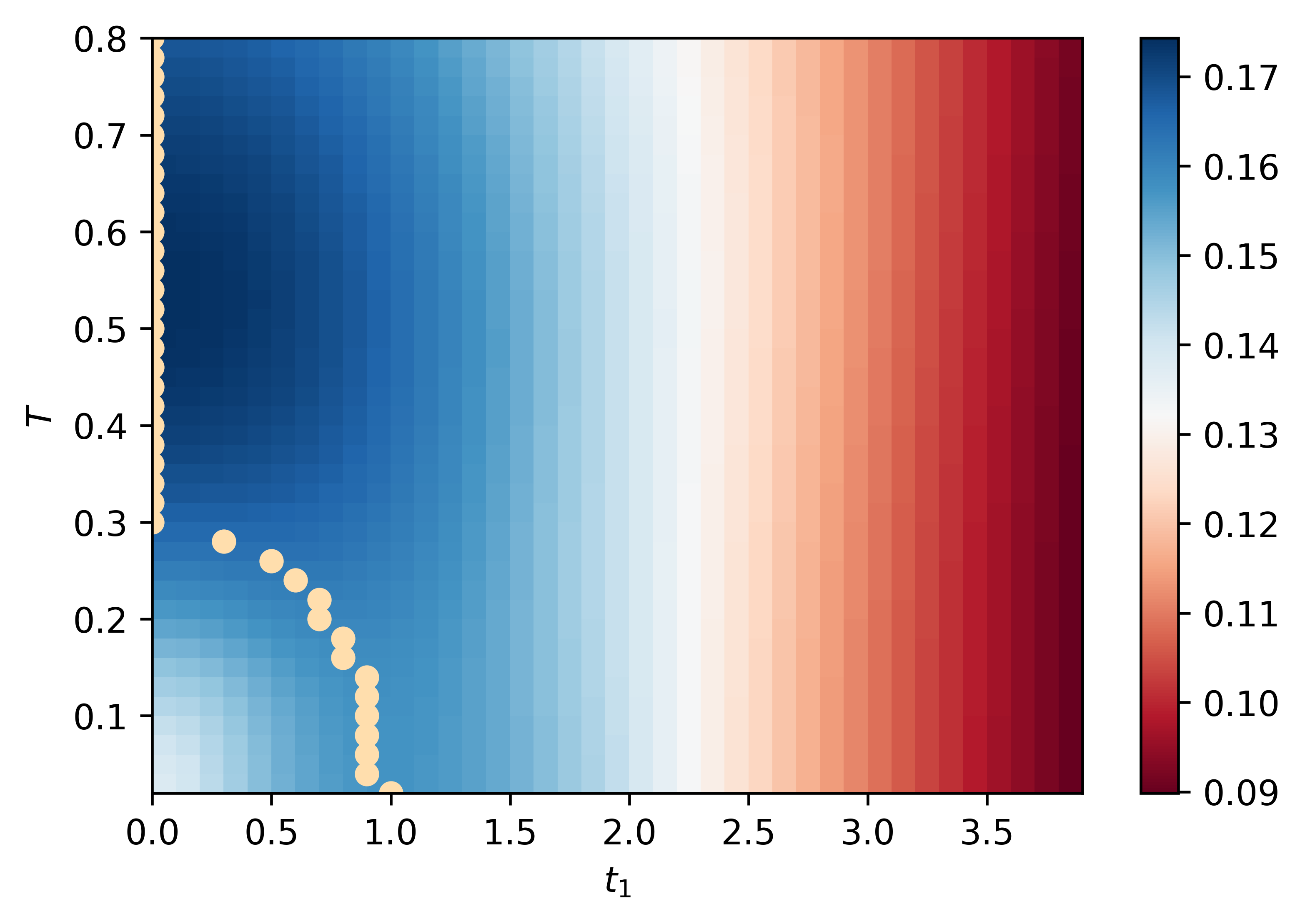}
\caption{$|\sigma_{H}|$ is plotted (in units of $e^2/h$) as a function of   $t_1$ and $T$ for $n=1$ mWSMs in the type-II phase at tilt $t_C = 1.5$. Parameters for the density plot are: $t_0 = 1, \mu = 0, t_z = 1, t = 1$. The Off-white circles represent the maximum value of $\sigma_H$ at each $T$. Temperatures at which circles migrate to the boundary at $t_1 = 0$ represent cases where there are no peaks in the AHE curve when varied with $t_1$.}
\label{density1}
\end{figure}

We focus on the AHE in the type-II mWSM phase, as shown in figure \ref{Figs4} (a). We note that at small values of $t_1$, the AHE shows non-monotonic variation with temperature and this effect is highlighted in Fig. \ref{Figs4} (c). Also, the $|\sigma_H|$ develops a peak for small temperatures, while for larger temperatures, no such peaks exists. Motivated by our results in the CME case, we investigate the DOS as a function of $t_1$ for this setup (see appendix). We find that no corresponding peaks exist in the DOS, thereby ruling it out as a potential explanation for the AHE maximas. We present an analysis of the origin of the peaks in the following paragraph.   

It is well known that the type-II phase hosts the unusual co-existence of electron-hole pockets and this leads to a unique distribution of available states close to the Fermi level, different from the type-I phase. So, certain thermal transitions (characterized by the temperature $T$) are preferred, and this preference (distinct from the type-I phase) changes when the electronic structure is altered by varying $t_1$. The optimal temperatures for maximal scattering processes are represented by the peaks in Fig. \ref{Figs4} (c). Additionally, these peaks occur because of difference in behavior of $|\sigma_H|$ between small and large values of $t_1$. For large $t_1$, the AHE should decrease since the energy is dominated by $t_1$ in the type-II phase. In fact, taking the $t_1 \rightarrow \infty$ limit leads to a vanishing AHE in Eqn. \ref{halle}. The type-II phase $|\sigma_H|$ peaks are then the result of a cross-over between these two regimes. At sufficiently high temperatures, the most probable thermal excitations are physically impossible since the energy of the system is bounded, and this leads to the absence of AHE peaks for high temperatures.

\begin{figure}
\centering
\subfigure[]{\includegraphics[scale=.55]{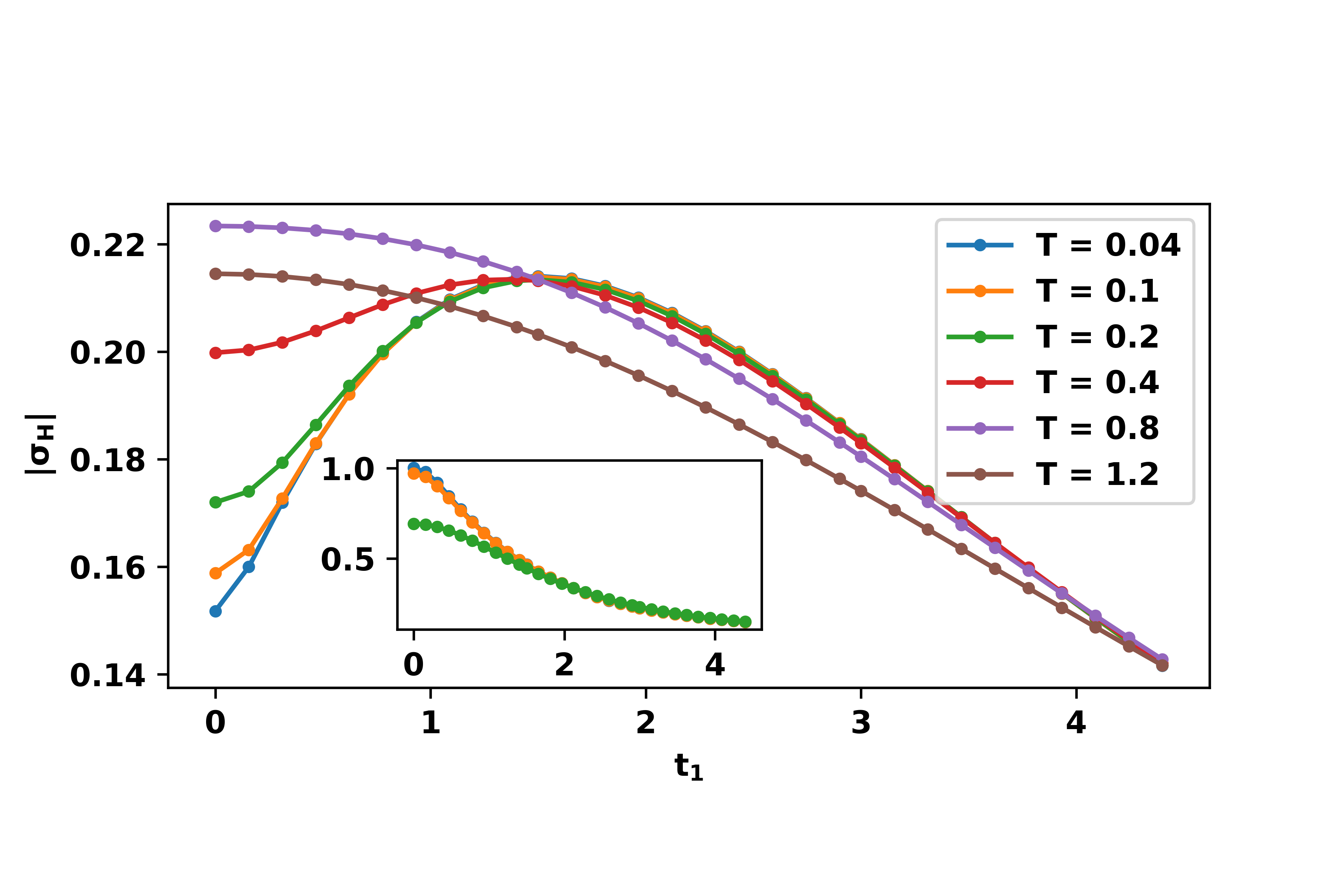}}
\subfigure[]{\includegraphics[scale=.55]{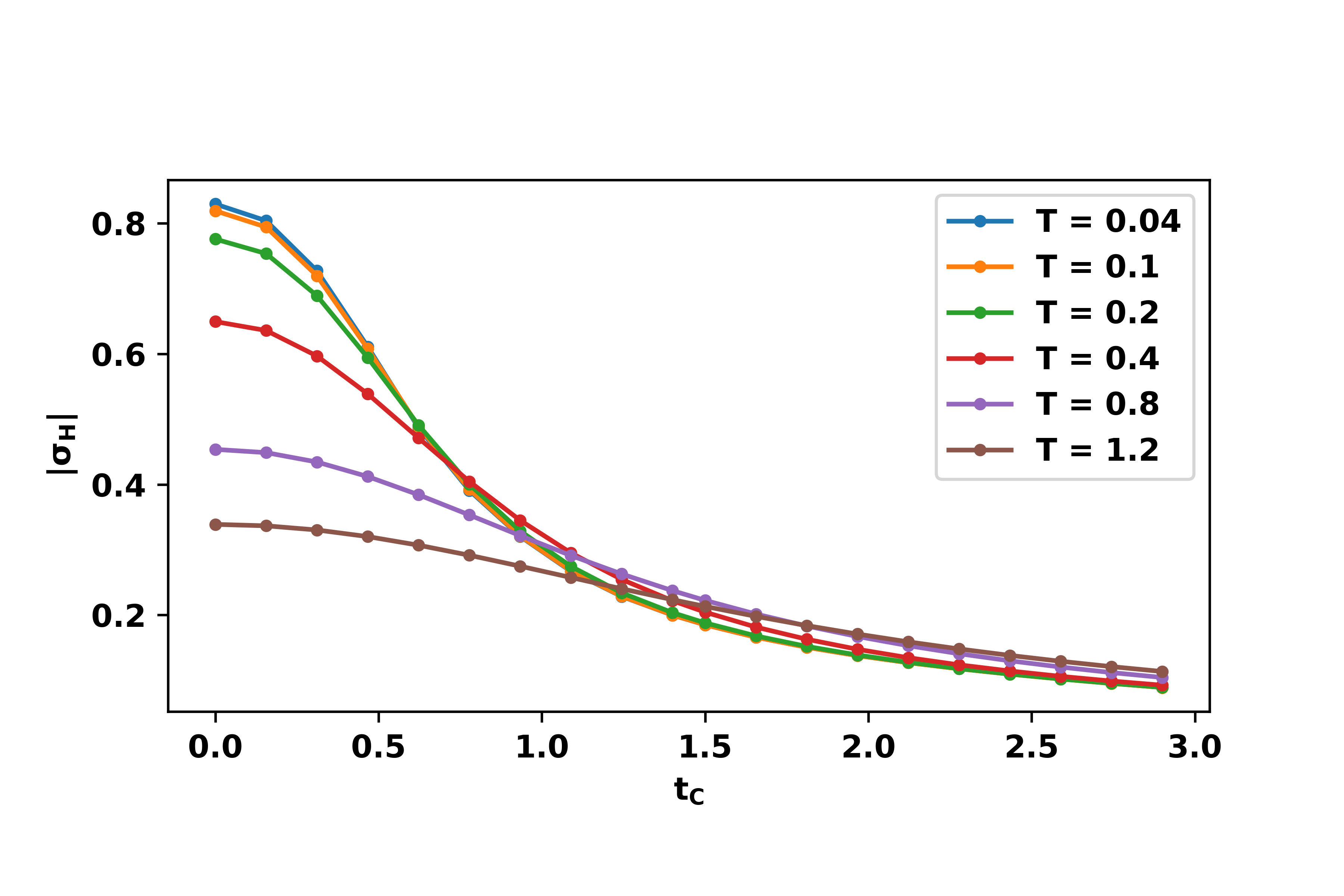}}
\subfigure[]{\includegraphics[scale=.55]{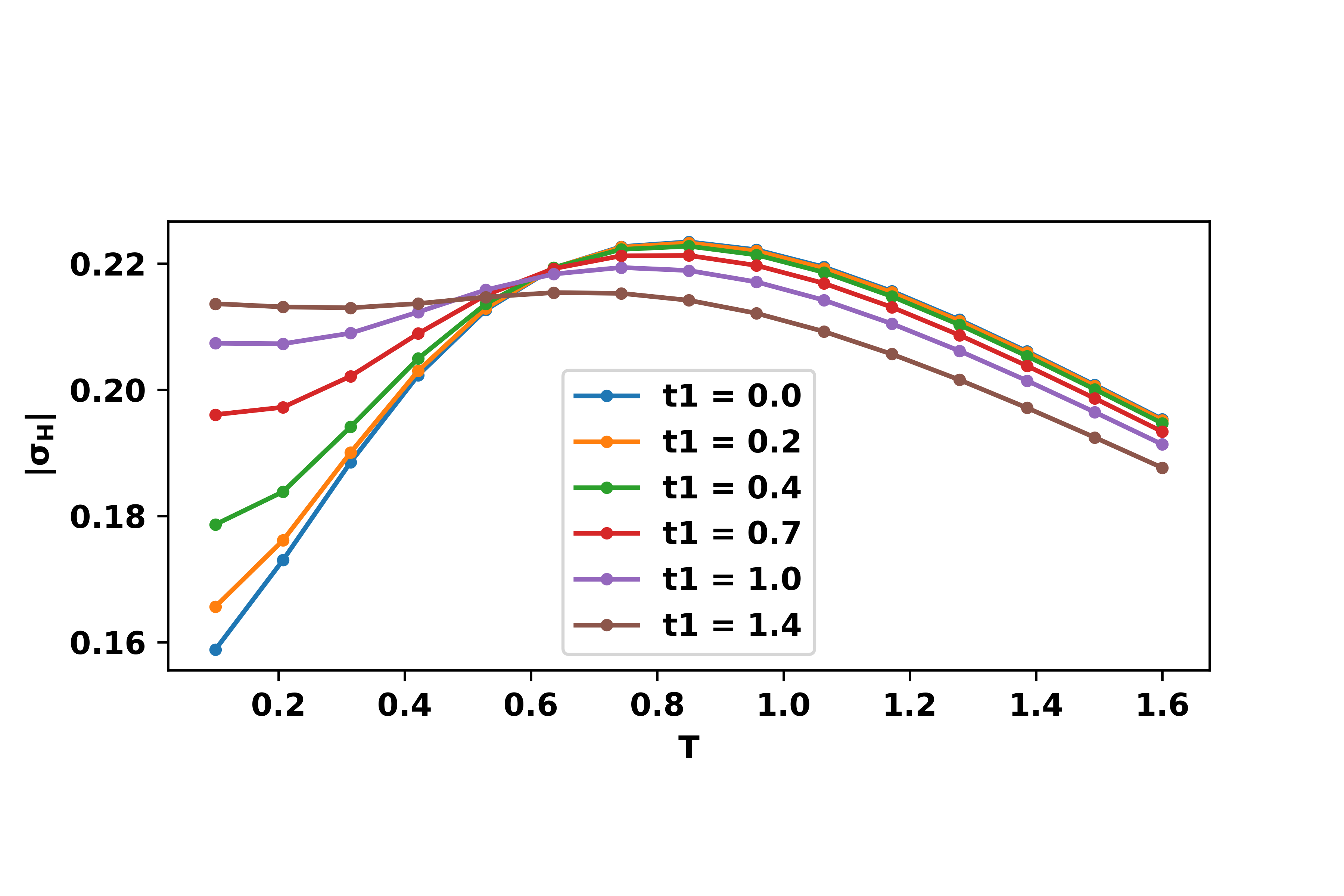}}
\caption{(a) The variation of $|\sigma_{H}|$ is plotted (in units of $e^2/h$) with $t_1$ for $n=2$ mWSMs, with the type-II phase shown in the mainframe ($t_C = 1.5$) and the type-I phase in the inset ($t_C = 0$). (b) The variation of $|\sigma_{H}|$ is plotted with $t_C$ for $n=2$ mWSMs ($t_1 = 0.5$). (c) The variation of $|\sigma_H|$ is plotted as a function of temperature for different values of $t_1$ for the $n=2$ mWSM with $t_C = 1.5$. For small values of $t_1$, the AHE shows a peak as a function of temperature. Parameters common to all plots are: $t_0 = 0.25, \mu = 0, t_z = 0.5, t = 0.5$.}
\label{Figs5}
\end{figure}

Additionally, in Fig. \ref{density1} we present the AHE as a function of $T$ and $t_1$, where maximum value of $|\sigma_H|$ at each temperature is represented by the Off-white circles. For small temperatures, we see the peaks in the AHE distribution, and these maximas migrate to lower values of $t_1$ with increasing temperature. At $T \geq 3$, the Off-white circles move to the boundary at $t_1 = 0$ indicating that there are no peaks in the distribution.

Fig. \ref{Figs4} (b) shows the variation of $|\sigma_H|$ as a function of tilt. Since a finite value of $t_1$ induces an effective chemical potential (by splitting the nodes in energy space), the AHE suffers Fermi-surface corrections in the type-I phase. Thermal effects promote a higher value of $|\sigma_H|$ at small values of $t_1$, though by contrast, they inhibit the AHE as a function tilt. On examining eqn.(\ref{halle}), we find that $|\sigma_H|$ should exponentially decay as $t_C \rightarrow \infty$. This behavior is clearly evident as a function of $t_C$ [Fig. \ref{Figs4} (b)] for our choice of parameters, while it is somewhat suppressed as a function of $t_1$ [Fig. \ref{Figs4} (a)]. The fact that tilting seems to have a more dominant effect on the behavior of the conductivity coefficients as compared to the energy separation of the Weyl nodes is reminiscent of the CME case. While it appears that the value of $|\sigma_H|$ becomes universal around the Lifshitz transition for small $T$ [see Fig. \ref{Figs4} (b)], this feature ceases to exist for $T = 0.8, 1.2$. 

\begin{figure}
\centering
\subfigure[]{\includegraphics[scale=.55]{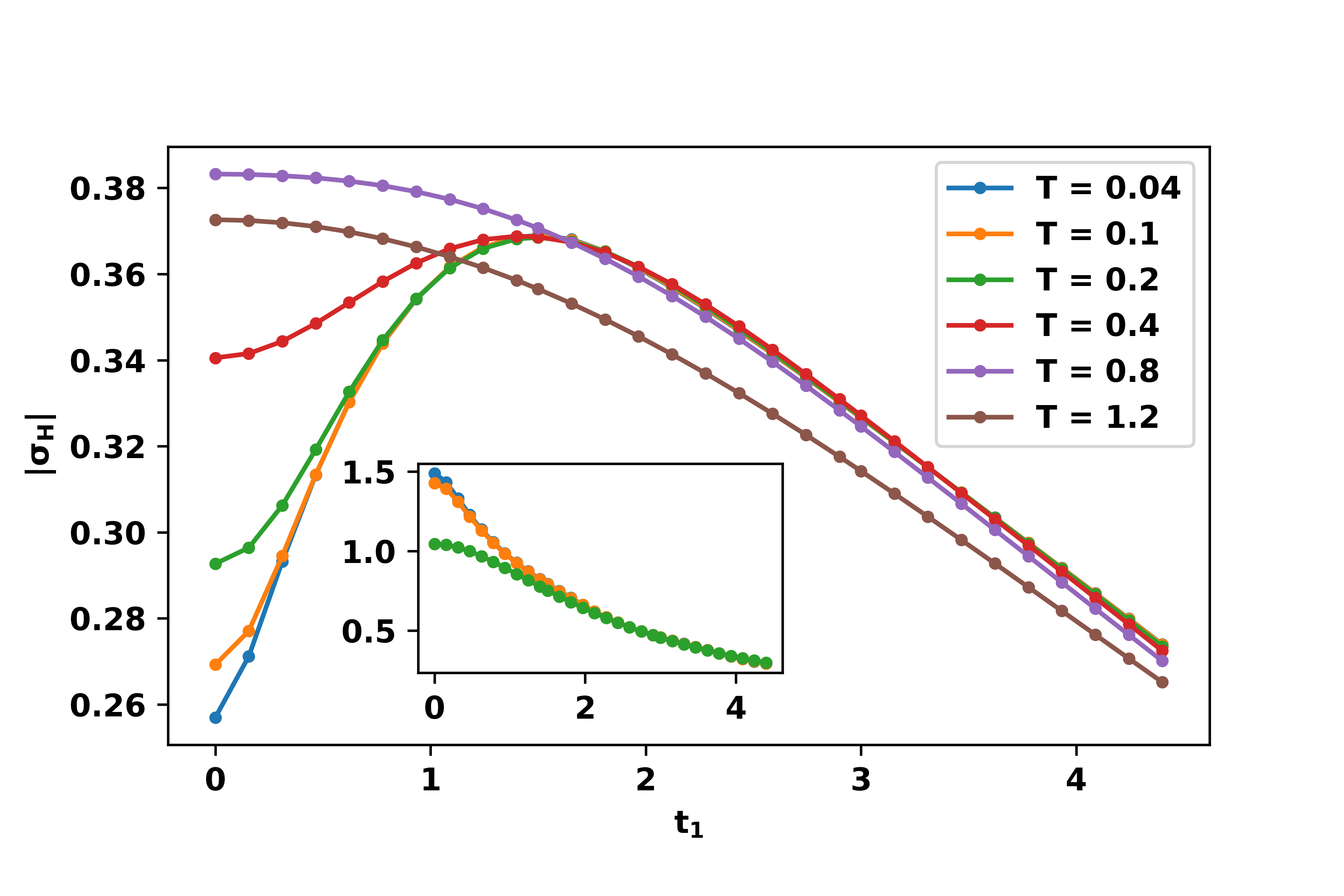}}
\subfigure[]{\includegraphics[scale=.55]{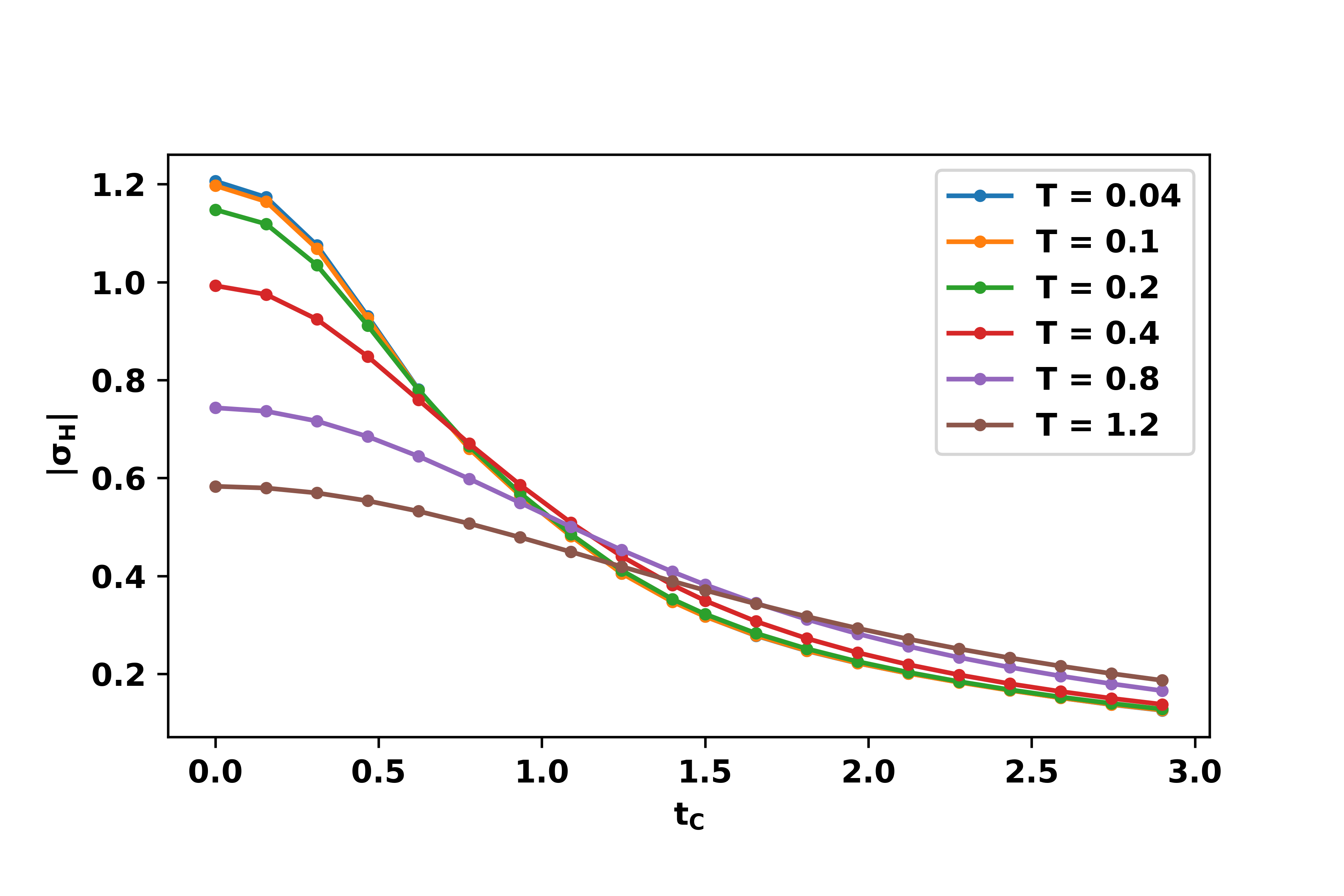}}
\subfigure[]{\includegraphics[scale=.55]{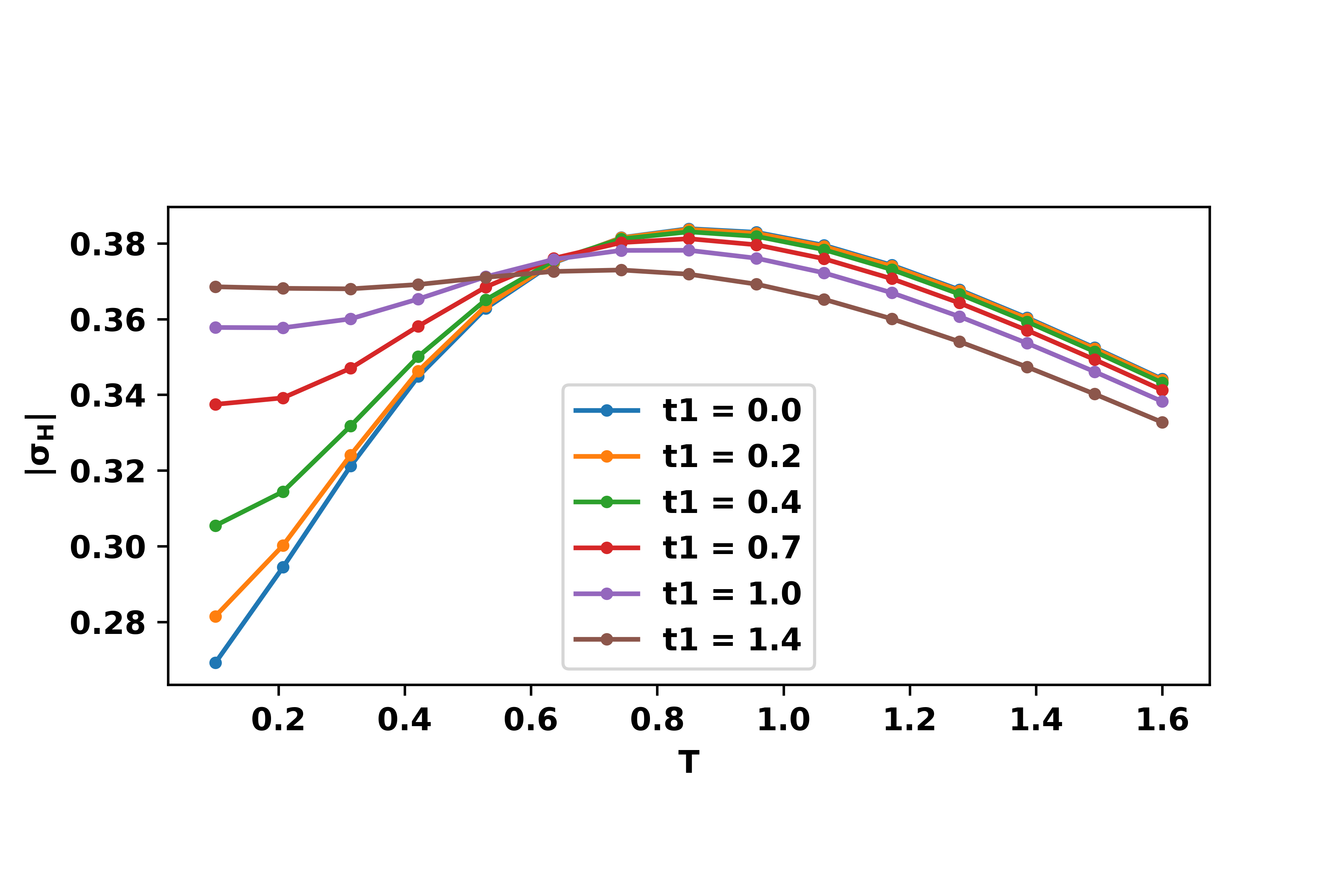}}
\caption{(a) The variation of $|\sigma_{H}|$ is plotted (in units of $e^2/h$) with $t_1$ for $n=3$ mWSMs, with the type-II phase shown in the mainframe ($t_C = 1.5$) and the type-I phase in the inset ($t_C = 0$). (b) The variation of $|\sigma_{H}|$ is plotted with $t_C$ for $n=3$ mWSMs ($t_1 = 0.5$). (c) The variation of $|\sigma_H|$ is plotted as a function of temperature for different values of $t_1$ for the $n=3$ mWSM with $t_C = 1.5$. For small values of $t_1$, the AHE shows a peak as a function of temperature. Parameters common to all plots are: $t_0 = 0.25, \mu = 0, t_z = 0.5, t = 0.5$.}
\label{Figs6}
\end{figure}

The AHE for $n = 2$ and $n = 3$ mWSMs as functions of $t_1$, $t_C$, and $T$, are shown in Fig. \ref{Figs5} [(a), (b) \& (c)] and Fig. \ref{Figs6} [(a), (b) \& (c)], respectively. Every single one of the features discussed in the context of the $n=1$ mWSMs generalize well to the higher monopole charge cases. As has been the norm throughout this work, we present the type-I results as a function of $t_1$ in the inset of Figs. \ref{Figs5} (a) and \ref{Figs6} (a), and the values of $|\sigma_H| = n\frac{e^2 Q}{2\pi \hbar}$ hold for $n=2, 3$. The peak in the plot of $|\sigma_H|$ vs $t_1$ in the type-II phase is robust to monopole charge and this feature is generated by the interplay between the tilt and the CME generating node separation parameter $t_1$. The trend of an enhanced AHE for small $t_1$ as a function of increasing temperature is also observed, as is the decrease in $|\sigma_H|$ for larger values of $t_1$. The plots of anomalous Hall conductivity as a function of $t_C$ are shown in figures \ref{Figs5} (b), \ref{Figs6} (b) for $n=2, 3$ respectively. 

While having control over material parameters, is in general, tricky, we point out a few ways to test our claims using recent advances. For example, strain causes changes in locations in the Brillouin zone, leading to varying tilt and energy separated Weyl nodes. This technique, which relies on the spatial and or temporal component of the elastic gauge field, may be used to generate a CME signal \cite{Prop1, Prop2, CME6}. Another line of approach would be to gain a controllable handle on tilt by employing periodic driving which has been reported to produce type-I and type-II WSMs \cite{Prop4}. These techniques can be utilized to probe the features of $\alpha$ and $|\sigma_H|$ described in this manuscript.


\section{Conclusion} 

In summary, we have systematically studied the chiral magnetic effect and the anomalous Hall effect in lattice models of tilted multi-Weyl semimetals for both phases and at different temperatures. This study is motivated by the dearth of literature regarding the nature of the chiral magnetic effect and anomalous Hall effect in the type-II phase of mWSMs, and its variation as a function of tilt. We find that the CME is mediated by the Berry curvature for $n=2,3$ mWSMs similar to the already established $n=1$ case \cite{CME1, CME2}. We have established characteristic features of the CME as a function of tilt in both phases which we believe are qualitatively robust. We find that $|\alpha|$ increases as a function of tilt in the type-I phase and decreases in the type-II phase with universal behavior for all three values of monopole charge. The phase boundary is marked by a peak which we believe to be a signature of the Lifshitz transition. We demonstrate the origin of the peak in terms of the density of states for $n=1$ mWSMs. As a function of $t_1$ (the energy split between the Weyl points) in the type-II phase, the CME appears to be monotonically increasing. While for smaller temperatures the CME grows rapidly, at larger temperatures this phenomenon is mildly suppressed by thermal fluctuations. The low temperature $t_C$ and $t_1$ dependent variations of the CME constitute the principal results of this manuscript.

As far as the anomalous Hall effect is concerned, we observe that $|\sigma_H|$ is a decreasing function of tilt in the type-II phase. In the type-II phase, $|\sigma_H|$ increases as a function of $t_1$ for small $t_1$, peaks, and then decays for large $t_1$. Similar to the case of the CME these properties are universal to all monopole charges. Lower temperatures appear to wash away the peak in the $|\sigma_H|$, and we note that thermalization enhances the AHE for small values of $t_1$. For larger values of $t_1$, $|\sigma_H|$ becomes temperature independent near the Lifshitz transition for small values of $T$, while this observation breaks down for sufficiently large temperatures. We believe that these results are robust since the lattice models don't suffer the divergencies of the more conventionally explored minimal models.

{\textit {\textbf{Acknowledgement}:}} A. M. would like to acknowledge that this work is supported in part by National Science Foundation grant number NSF-DMR 1855111. S. C. acknowledges the support of Physics and Applied Mathematics Unit, Indian Statistical Institute.\\

\appendix

\section{Minimal model Hamiltonian and Dispersion}

In this section we would like to analyze the uniform limit CME in the minimal model, as computed in \cite{CME0,CME1,ZB1}. We begin our journey by reviewing the minimal model Hamiltonian for a TRS broken mWSM \cite{AMBB,AMBBTN} \\

\begin{align}
H_{n}^{s}=\hbar C_{s}(k_z-sQ)+s\hbar \beta_n \bm\sigma\cdot{\bm n}_p,
\label{1}
\end{align}

where $s=\pm$ characterizes the Weyl point (WP), $C_s$ is the tilt parameter, which can be different for each node, in principle. Here, ${\bm n}_p=\frac{1}{\hbar} \left[p_{\perp}^n\cos(n\phi_p), p_{\perp}^n\sin(n\phi_p), \frac{v(p_z- s\hbar Q)}{\beta_n}  \right]$, $p_{\perp} =\sqrt{p_x^2+p_y^2}$, $\bm \sigma$ is the vectorized Pauli matrix, $v$ denotes the Fermi velocity in the absence of tilt, and $n$ is the monopole charge. This Hamiltonian has mW nodes separated by $2Q$ along $\bm e_z$, which is the unit vector along the z-direction in momentum space. $\beta_n$ constitutes the dimensionally consistent generalization of the Fermi velocity in the $k_x-k_y$ plane. \\

We proceed by introducing a magnetic field perpendicular to the $x-y$ plane in the Landau gauge: $\bm A=x B {\hat{\bm y}}$ such that $\bm B=\bm {\nabla} \times \bm A=-B {\hat{\bm z}}.$ The Hamiltonian is presented in the compact matrix form:

\begin{equation}\label{ham6} 
H^n_{s,B} = \left[ \begin{array}{cc}  (C_s + s v) z  &  i^n s \frac{\beta_n}{\ell_B^n}(\sqrt{2} a^\dagger)^n \\ \\
  -  i^n s \frac{\beta_n}{\ell_B^n}( \sqrt{2} a)^n    & (C_s - s v) z  
  \end{array} \right] ,
\end{equation}

with the introduction of the ladder operators $a(a^{\dagger})$, following the Pierel's substitution $p_i \rightarrow p_i - eA_i$, and setting $c = 1$, $\hbar =1$, $z = k_z -sQ$. 

\begin{align}\label{en2}
E_N^{t,s} &= C_s z+t v \sqrt{z^2+ {^NP_n} \Omega^2}, \nonumber \\
\end{align}

where, $\Omega$ represents the LL spacing. Note that that the $N$ degenerate ground states are chiral and host only the states corresponding to $s \cdot t  =-1$. \\

We look to extend our tilted mWSM model to reproduce the CME results of \cite{CME1}. We note that the model Hamiltonian used in \cite{ZB1} is 

\begin{equation}\label{zb1}
H = \frac{\omega_B}{\sqrt{2}} (\sigma^+ a + \sigma^- a^{\dagger})\tau^z +\tau^z \sigma^z k_z + \tau^z b_0 -\Delta \tau_x. 
\end{equation}

Let's understand this model briefly. If we set $\Delta = 0$, we see that this model is two copies of an un-tilted Weyl node in a magnetic field with Hamiltonian $H_0 = \frac{\omega_B}{\sqrt{2}} (\sigma^+ a + \sigma^- a^{\dagger}) + \sigma^z k_z$, at different energies $\pm b_0$. The $\Delta$ term is a node hybridization term introduced by \cite{ZB1}, and it gaps out the WPs at $\Delta \neq 0$. The $\sigma$ matrices represent the valley degrees of freedom, and the $\tau$ matrices denote the node d.o.f. Motivated by this, we first rewrite the Hamiltonian in Eqn.(\ref{ham6}) in a more compact form:

\begin{equation}
H_{s,B}^n = s\Omega [\sigma^+ a^n + \sigma^- (a^{\dagger})^n] + sv\sigma^z k_z + C_s k_z. 
\end{equation}

We have removed the node separation in momentum space $Q$ which is unrelated to the CME. We can now restrict to the tilt symmetric case which is relevant to the lattice model analysis that follows: $C_+ = -C_- = C$, or, $C_s = sC$. With this, we can recast our mWSM Hamiltonian in the form presented in Eqn.(\ref{zb1}) as shown below:

\begin{align}
H = & \Omega[\sigma^+ a^n + \sigma^- (a^{\dagger})^n]\tau^z + v\tau^z \sigma^z k_z +  \tau^z (b_0 + C k_z) \nonumber \\ & -\Delta \tau_x. 
\end{align}

One can check that for $\Delta = 0$ and $b_0 =0$, one recovers the Hamiltonian in Eqn.(\ref{ham6}). The dispersion for such Hamiltonian would be given by

\begin{align}
E_N^{s,\gamma} &= s\sqrt{\left [\gamma (b_0 + C k_z)+ \sqrt{v^2k_z^2+ {^NP_n} \Omega^2} \right]^2 + \Delta^2}.
\end{align}

One can then consider the $\Delta \rightarrow 0$ limit and set $C = 0$ and $n=1$ to recover the dispersion in \cite{ZB1}. $\gamma = \pm$ indicates the node d.o.f as can be seen by its effect on $b_0$ and $C$, and $s=\pm$ represents the valley d.o.f. In the case $N < n$, we have the $n$ degenerate chiral ground states which are characterized by the dispersions: 

\begin{align}
E_0^{\gamma} &= \gamma\sqrt{\left [(v-C) k_z - b_0\right]^2 + \Delta^2}.
\end{align}

Again, this can be checked with the results in \cite{ZB1} by setting $C=0$.

\section{Chiral Magnetic Effect in the minimal model: Uniform Limit}

We would like to motivate the dependence of the CME on tilt by examining the consequence of a minimal model calculation. Following \cite{CME1,CME0,ZB1}, we get that the equilibrium current in response to the applied magnetic field can be calculated by the following formula:

\begin{widetext}
\begin{align}\label{e5}
    J_z &= -\frac{e^2B}{2\pi \hbar^2} \int_{\Lambda}^{-\Lambda} \frac{dk_z}{2\pi} \frac{d}{dz} \left[ n E_0^{-} + \sum_{m=n}^{N_{max}} \sum_{\gamma = \pm} E_m^{-,\gamma} \right] 
\end{align}
\end{widetext}

$N_{max}$ is a LL cutoff as introduced in \cite{CME0}, and has been used previous to our work on Landau mWSMs \cite{AMBB}.  In fact, as pointed out in \cite{CME1}, the correct choice of limits decides whether we are in the static limit or uniform limit of the CME. The expression was originally derived in the context of gluons in \cite{CME0} by defining an appropriate thermodynamic potential. While it is a useful exercise to understand this derivation, we focus our attention towards evaluating this object for our model. Since the expression inside the integral is a total derivative, the answer is obtained as

\begin{widetext}
\begin{align}\label{e6}
    \alpha = -\frac{J_z}{B} &= \frac{e^2}{4\pi^2 \hbar^2} \left[ n E_0^{-} + \sum_{m=n}^{N_{max}} \sum_{\gamma = \pm} E_m^{-,\gamma} \right]_{\Lambda}^{\Lambda}.
\end{align}
\end{widetext}

\subsection{The CME in type-I mWSMs}

With this, we proceed to evaluate the CME parameter $\alpha$ in the type-I and type-II regime. For $N>n$, one can see that when $C=0$ the second term in Eqn.(\ref{e6}) is even in $k_z$, and hence vanishes identically. However, when $C \neq 0$ we may get a non-trivial contribution. In the type-I phase and away from the Lifshitz transition, we have that: $C \ll v$. We evaluate the limit:

\begin{widetext}
\begin{align}
    \lim_{\Lambda \rightarrow \infty} E_m^{-,\gamma} (\Lambda) - E_m^{-,\gamma} (-\Lambda)  = & -\sqrt{\left [\gamma (b_0 + C \Lambda)+ \sqrt{v^2 \Lambda^2+ {^NP_n} \Omega^2} \right]^2 + \Delta^2} \nonumber \\ & + \sqrt{\left [\gamma (b_0 - C \Lambda)+ \sqrt{v^2 \Lambda^2+ {^NP_n} \Omega^2} \right]^2 + \Delta^2} \nonumber \\ = & {\Bigg |} \gamma b_0 +(v-\alpha C) \Lambda {\Bigg |} - {\Bigg |} \gamma b_0 +(\gamma C+v) \Lambda {\Bigg |} \nonumber \nonumber \\ = &  -2\gamma C \Lambda
\end{align}
\end{widetext}

This implies that the sum in the second term in Eqn.(\ref{e6}) vanishes: $-2C\Lambda \sum_{\gamma = \pm} \gamma = 0$. The CME for this mWSM receives no correction from the excited states! We evaluate the same limit for the ground state, which is the first term in Eqn.(\ref{e6}):

\begin{widetext}
\begin{align}
    \lim_{\Lambda \rightarrow \infty} E_0^{-} (\Lambda) - E_0^{-} (-\Lambda)  = & -\sqrt{\left [  (v-C) \Lambda -b_0 \right]^2 + \Delta^2} \nonumber + \sqrt{\left [ -(v-C) \Lambda - b_0 \right]^2 + \Delta^2} \nonumber \\ = & {\Bigg |} -(v-C) \Lambda - b_0{\Bigg |} - {\Bigg |} (v-C) \Lambda - b_0 {\Bigg |} \nonumber \nonumber \\ = &  (v-C)\Lambda + b_0 - (v-C)\Lambda + b_ 0 \nonumber \\  = & 2b_0.
\end{align}
\end{widetext}

Thus, we obtain the CME parameter in the type-I tilted mWSM phase as 

\begin{equation}
    \alpha = n\frac{e^2}{2\pi^2 \hbar^2}b_0,
\end{equation}

where the monopole charge enhances the CME effect and tilt does not affect the results. 

\subsection{The CME in type-II mWSMs}

Now we can consider what happens in the type-II phase of the tilted mWSM: $C \gg v$. Here, we need to reevaluate the two limits with the constraint on tilt: 

\begin{widetext}
\begin{align}\label{e9}
    \lim_{\Lambda \rightarrow \infty} E_m^{-,\gamma} (\Lambda) - E_m^{-,\gamma} (-\Lambda)  = & -\sqrt{\left [\gamma (b_0 + C \Lambda)+ \sqrt{v^2 \Lambda^2+ {^NP_n} \Omega^2} \right]^2 + \Delta^2} \nonumber \\ & + \sqrt{\left [\gamma (b_0 - C \Lambda)+ \sqrt{v^2 \Lambda^2+ {^NP_n} \Omega^2} \right]^2 + \Delta^2} \nonumber \\ = & {\Bigg |} \gamma b_0 +(v-\alpha C) \Lambda {\Bigg |} - {\Bigg |} \gamma b_0 +(\gamma C+v) \Lambda {\Bigg |} \nonumber \nonumber \\ = &  
    C \Lambda - \gamma v\Lambda - \gamma b_0 - C \Lambda - \gamma v\Lambda + \gamma b_0 \nonumber \\ = &
    -2\gamma v\Lambda 
\end{align}
\end{widetext}

The sum over gamma kills this term. The zero-mode limit is evaluated similarly to $-n\frac{e^2}{2\pi^2 \hbar^2}b_0$. We combine these results to get the CME in the type-II phase as:

\begin{equation}
    \alpha = -n\frac{e^2}{2\pi^2 \hbar^2}b_0,
\end{equation}

It appears that the Lifshitz transition affects the CME in the minimal model. So, it suggests that the CME may be influenced by tilt in at least one phase. However, minimal models are plagued by unphysicalities like infinite electron-hole pockets and results derived using it are not always trustworthy. In what follows, we examine the CME using a set of lattice models in the manuscript. \\

\section{Berry curvature}

To evaluate the integrals in (\ref{halle}) and (\ref{cmef}) numerically we write the explicit expressions of the Berry curvature for the $n=1$ and $n=2$ cases of mWSMs. The explicit expression of the Berry curvature  for $n=3$ is very complicated and hence  not included  here. 

For $n=1$, components of the Berry curvature are given by

\begin{widetext}
\begin{eqnarray}
\Omega_{\pm}^z &=& \pm\frac{t^2 \cos k_x \cos k_y \{t_0 (2 - \cos k_x - \cos k_y) + t_z \cos k_z\} - 
 t^2 t_0 \cos k_y \sin^2 k_x - t^2 t_0 \cos k_x \sin^2 k_y}{2 \{(t_0 (2 - \cos k_x - \cos k_y) + t_z \cos k_z)^2 + 
   t^2 \sin^2 k_x + t^2 \sin^2 k_y\}^{3/2}}\nonumber \\
\Omega_{\pm}^y &=& \pm\frac{t^2 t_z \cos k_x \sin k_y \sin k_z}{2 \{(t_0 (2 - \cos k_x - \cos k_y) + t_z \cos k_z)^2 + 
   t^2 \sin^2 k_x + t^2 \sin^2 k_y\}^{3/2}}\nonumber \\
   \Omega_{\pm}^x &=& \pm\frac{t^2 t_z \cos k_y \sin k_x \sin k_z}{2 \{(t_0 (2 - \cos k_x - \cos k_y) + t_z \cos k_z)^2 + 
   t^2 \sin^2 k_x + t^2 \sin^2 k_y\}^{3/2}} 
\end{eqnarray}
\end{widetext}

For $n=2$, components of Berry curvature are:

\begin{widetext}
 \begin{eqnarray}
 \Omega_{\pm}^z=\{(t_0 (6 - 4 \cos k_x + \cos 2k_x - 4 \cos k_y + \cos 2k_y) + \nonumber\\
      t_z \cos k_z) (-2t^2 \cos k_y \sin k_x^2 - 2t^2 \cos k_x \sin k_y^2) + \nonumber\\
   2t \sin k_x \sin k_y (8 t t_0 \sin k_x \sin k_y - 2 t t_0 \sin 2k_x \sin k_y - \nonumber\\
      2 t t_0 \sin k_x \sin 2k_y) + t (\cos k_x - \cos k_y) (-8 t t_0 \cos k_y \sin k_x^2 + \nonumber\\
      4 t t_0 \cos k_y \sin k_x \sin 2k_x + 8 t t_0 \cos k_x \sin k_y^2 - 4 t t_0 \cos k_x \sin k_y \sin 2k_y\}/\nonumber\\
      \{2 (t^2 (\cos k_x - \cos k_y)^2 + (t_0 (6 - 4 \cos k_x + \cos 2k_x - 4 \cos k_y + \nonumber\\
          \cos 2k_y) + t_z \cos k_z)^2 + 4t^2 \sin k_x^2 \sin k_y^2)^{3/2}\}
          \end{eqnarray}
  \begin{eqnarray}   
 \Omega_{\pm}^y = \pm\{2t^2 t_z \cos k_x (\cos k_x - \cos k_y) \sin k_y \sin k_z + 2t^2 t_z \sin k_x^2 \sin k_y \sin k_z\}/\nonumber\\
 \{2 (t^2 (\cos k_x - \cos k_y)^2 + \nonumber\\
 (t_0 (6 - 4 \cos k_x + \cos 2k_x - 4 \cos k_y + 
        \cos 2k_y) + t_z \cos k_z)^2 +\nonumber\\ 4t^2 \sin k_x^2 \sin k_y^2)^{3/2}\} \nonumber\\
        \end{eqnarray}
\begin{eqnarray}
 \Omega_{\pm}^x=\pm\{-2t^2 t_z (\cos k_x - \cos k_y) \cos k_y \sin k_x \sin k_z + 
 2t^2 t_z \sin k_x \sin k_y^2 \sin k_z\}/\nonumber\\
 \{2 (t^2 (\cos k_x - 
      \cos k_y)^2 + (t_0 (6 - 4 \cos k_x + \cos 2k_x - 4 \cos k_y +\nonumber \\
        \cos 2k_y) + t_z \cos k_z)^2 + 4t^2 \sin k_x^2 \sin k_y^2)^{3/2}\} \nonumber\\
        \end{eqnarray}

\end{widetext}

The integrals are evaluated numerically with  the number of lattice cites  $N\geq 400 ^3$ 

\section{Density of States}

The DOS for type-II $n=1$ mWSM is presented below in Fig. \ref{DOSt1} as a function of $t_1$. This plot has no peak at $t_1 \sim 1$, where the AHE shows a peak. This clarifies that the peak in the AHE vs $t_1$ plot is uncorrelated with DOS.

\begin{figure}
\centering
\includegraphics[scale=0.6]{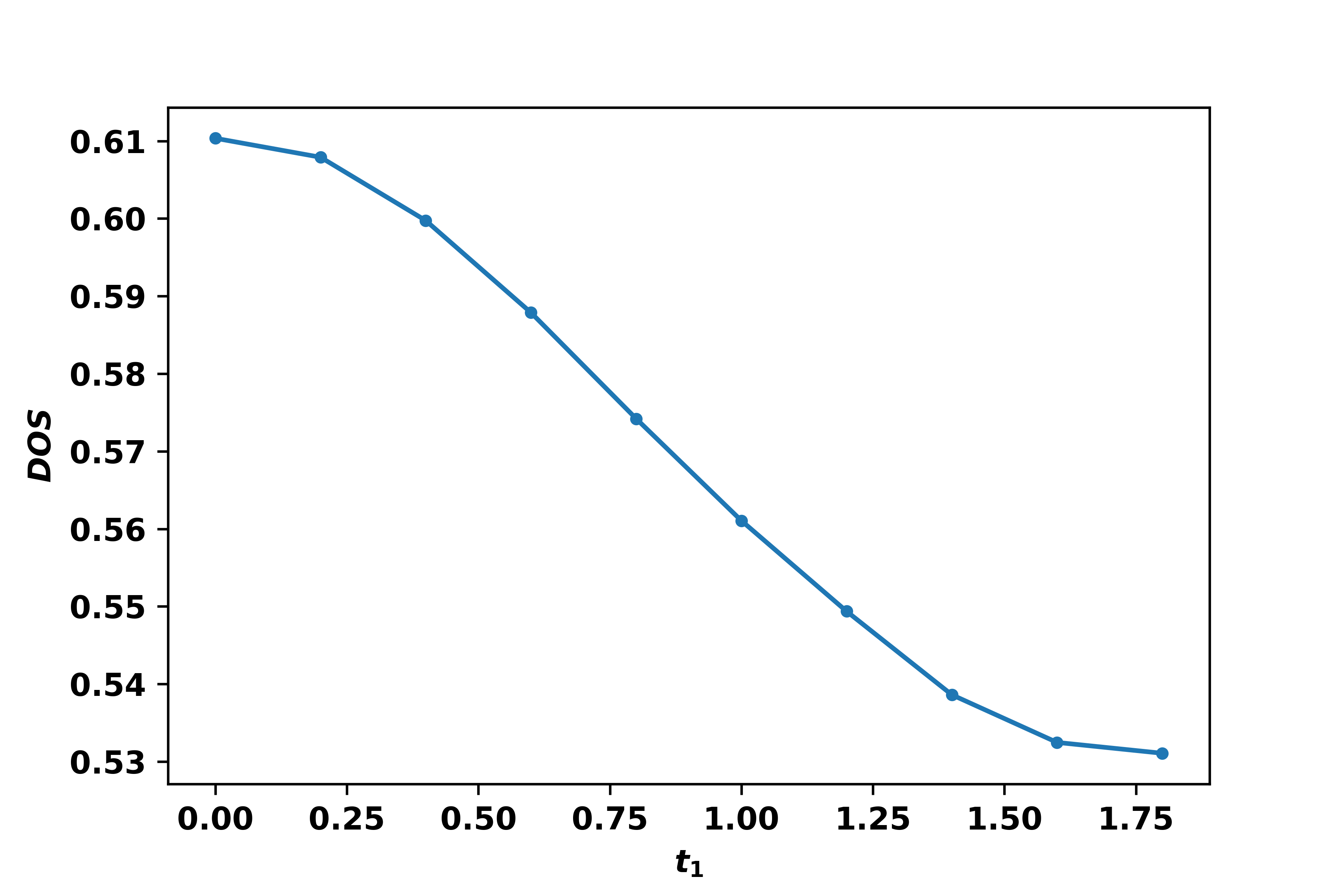}
\caption{The DOS is plotted with $t_1$ for $n=1$ mWSMs, with the type-II phase shown in the mainframe ($t_C = 1.5$). Parameters for the plot are: $t_0 = 1, \mu = 0, t_z = 1, t = 1$. No peak is observed in the DOS as a function of $t_1$ for the range sampled.}
\label{DOSt1}
\end{figure}

\end{document}